\documentclass[sigconf]{acmart}
\pdfoutput=1
\usepackage{algorithm}      % 算法浮动体环境
\usepackage{algpseudocode}

\usepackage{subfigure}
\usepackage{xcolor}
\usepackage{booktabs}
\usepackage{multirow}
\usepackage{hyperref}
\usepackage{footmisc} % 用于控制脚注格式
\usepackage{amsmath}
\usepackage{bm}
\AtBeginDocument{%
  }

\setcopyright{acmlicensed}
\copyrightyear{2018}
\acmYear{2018}
\acmDOI{XXXXXXX.XXXXXXX}
\acmConference[Conference acronym 'XX]{Make sure to enter the correct
  conference title from your rights confirmation email}{June 03--05,
  2018}{Woodstock, NY}
\acmISBN{978-1-4503-XXXX-X/2018/06}

\begin{document}

\title{When Relevance Meets Novelty: Dual-Stable Periodic Optimization for Serendipitous Recommendation}

%%Exploratory Recommendation
% \author{Author Name\authornotemark[1]}
% \authornotetext[1]{Note content}
\author{Hongxiang Lin}
% \authornotetext[1]{Equal contribution}
\authornotemark[1]

% \authornote{1}
% \authornote{These authors contributed equally.}
\email{linhongxiang02@meituan.com}
\affiliation{%
  \institution{Meituan}
  \city{Beijing}
  \country{China}
}

\author{Hao Guo}
\authornote{Equal contribution.}
\email{guohao15@meituan.com}

\affiliation{%
  \institution{Meituan}
  \city{Beijing}
  \country{China}
}

\author{Zeshun Li}
\authornote{This work was done during the internship at Meituan.}
\email{lzs23@mails.tsinghua.edu.cn}
\affiliation{%
  \institution{Tsinghua University}
  \city{Beijing}
  \country{China}
}
\author{Erpeng Xue}
\email{xueerpeng@meituan.com}
\affiliation{%
  \institution{Meituan}
  \city{Beijing}
  \country{China}
}
\author{Yongqian He}
\email{heyongqian@meituan.com}
\affiliation{%
  \institution{Meituan}
  \city{Beijing}
  \country{China}
}

\author{Zhaoyu Hu}
\email{huzhaoyu02@meituan.com}
\affiliation{%
  \institution{Meituan}
  \city{Beijing}
  \country{China}
}

\author{Lei Wang}
\email{wanglei46@meituan.com}
\authornote{Corresponding author.}
\affiliation{%
  \institution{Meituan}
  \city{Beijing}
  \country{China}
}

\author{Sheng Chen}
\email{chensheng19@meituan.com}
\affiliation{%
  \institution{Meituan}
  \city{Beijing}
  \country{China}
}
\author{Long Zeng}
\email{zenglong@sz.tsinghua.edu.cn}
\authornotemark[3]
\affiliation{%
  \institution{Tsinghua University}
  \city{Beijing}
  \country{China}
}
% \affiliation{%

\renewcommand{\shortauthors}{Lin et al.}

\begin{abstract}

Traditional recommendation systems tend to trap users in strong feedback loops by excessively pushing content aligned with their historical preferences, thereby limiting exploration opportunities and causing content fatigue. Although large language models (LLMs) demonstrate potential with their diverse content generation capabilities, existing LLM-enhanced dual-model frameworks face two major limitations: first, they overlook long-term preferences driven by group identity, leading to biased interest modeling; second, they suffer from static optimization flaws, as a one-time alignment process fails to leverage incremental user data for closed-loop optimization.
To address these challenges, we propose the Co-Evolutionary Alignment (CoEA) method. For interest modeling bias, we introduce Dual-Stable Interest Exploration (DSIE) module, jointly modeling long-term group identity and short-term individual interests through parallel processing of behavioral sequences. For static optimization limitations, we design a Periodic Collaborative Optimization (PCO) mechanism. This mechanism regularly conducts preference verification on incremental data using the Relevance LLM, then guides the Novelty LLM to perform fine-tuning based on the verification results, and subsequently feeds back the output of the continually fine-tuned Novelty LLM to the Relevance LLM for re-evaluation, thereby achieving a dynamic closed-loop optimization.
Extensive online and offline experiments verify the effectiveness of the CoEA model in serendipitous recommendation.
\end{abstract}


% \ccsdesc[500]{Do Not Use This Code~Your Paper}



\maketitle

\section{Introduction}
% 方法对比
\begin{figure}[t]
\centering
\includegraphics[width=0.8\linewidth]{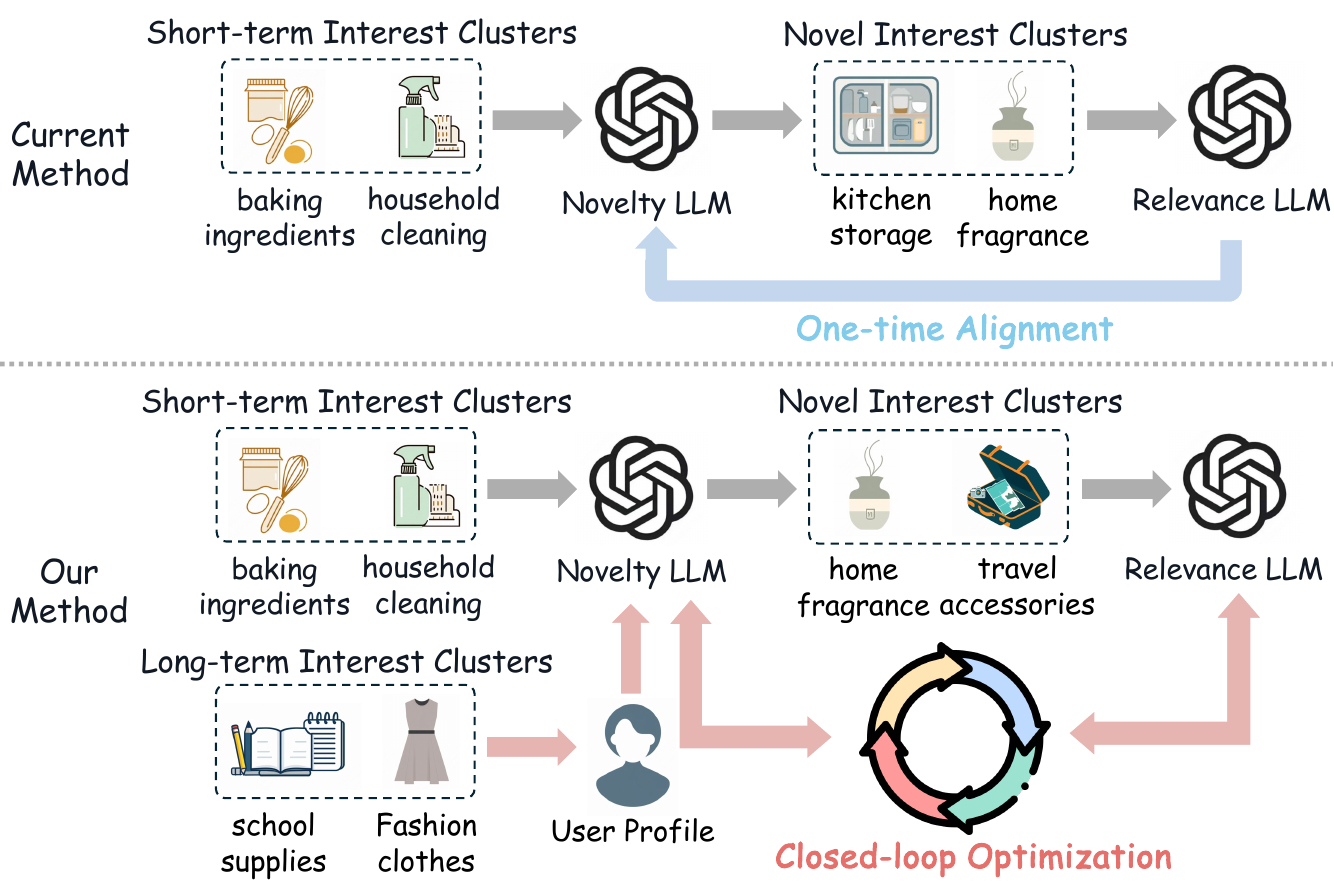} 
% \vspace{-cm}
\caption{Comparison between current and our methods.}
\label{intro}
\end{figure}
Recommendation systems aim to bridge the information gap between users and massive content by modeling user preferences. While traditional methods are effective, they tend to fall into the trap of a strong feedback loop \cite{mansoury2020feedback}, characterized by over-recommending content aligned with users' historical preferences. This not only limits users' opportunities to explore new interests but also easily triggers the phenomenon of content fatigue. Given the wide variety of interest types and the difficulty in assessing users' preferences for unfamiliar topics, the challenge of effectively presenting novel content to users is significantly heightened \cite{wu2024result}.

Large language models (LLMs) \cite{achiam2023gpt, guo2025deepseek, yang2025qwen3, chen2025minimax}, with their powerful reasoning capabilities and massive knowledge, can generate diverse recommended content. Existing methods \cite{chen2021exploration, lin2024rella, wang2024llms} reduce reliance on historical behaviors by introducing novel content to enhance long-term engagement. As shown in Figure \ref{intro}, current methods \cite{wang2025user, wang2025serendipitous,li2023hierarchical} adopt a hierarchical framework: using the \textit{Novelty LLM} to explore diversity and the \textit{Relevance LLM} to handle relevance, respectively. However, two key challenges remain:
\begin{itemize}
    \item \textbf{Interest Modeling Bias}. Existing methods adopt short-term interest sequences, making it difficult to capture the long-term stable preferences implied by users' group identities (e.g., "technology enthusiasts"). This causes novelty exploration to deviate from real needs, affecting recommendation acceptance and user satisfaction.
    \item \textbf{Dual-Model Collaboration Flaws}. After obtaining feedback from the \textit{Relevance LLM}, existing methods only align the \textit{Novelty LLM} once and generate a static database. They cannot continuously integrate dynamically updated user behavior information, which hinders closed-loop optimization based on incremental data.
\end{itemize}

To address the aforementioned challenges, we propose the \textbf{Co}-\textbf{E}volutionary \textbf{A}lignment method (CoEA), as shown in Figure \ref{intro}. To tackle the core issue of the lack of group identity in interest modeling, we design the Dual-Stable Interest Exploration (DSIE) module. This module achieves the collaborative modeling of group and individual interests by processing users' long-term and short-term behavioral sequences in parallel. Specifically, it conducts in-depth semantic modeling and clustering analysis on long-term behavioral sequences to extract user group Collaborative Semantic IDs (CSID) that can represent stable group preferences and their corresponding group portraits. For short-term behavioral sequences, it generates fine-grained sets of recent interest categories through the mapping mechanism. Thus, the DSIE module can simultaneously construct long-term and short-term user representations that integrate the stability of group identity and the timeliness of individual interests. 

To resolve the issue that static optimization of dual models fails to continuously integrate incremental data, we propose a Periodic Collaborative Optimization (PCO) mechanism. This mechanism constructs a bidirectional closed-loop iteration: the \textit{Novelty LLM} generates candidate interests based on dynamic user behavior data, while the \textit{Relevance LLM} performs preference verification and alignment. The evaluation signals output by the \textit{Relevance LLM} not only drive the incremental fine-tuning of the \textit{Novelty LLM}, enabling it to continuously adapt to preference evolution, but also incorporate a constraint loss to avoid catastrophic forgetting, effectively maintaining the model's knowledge retention and new interest mining capabilities. The output of the incrementally optimized \textit{Novelty LLM} is then used by the \textit{Relevance LLM} for re-scoring, thereby forming a sustainably optimizable collaborative cycle.

Our contributions are summarized as follows:
\begin{itemize}
    \item \textbf{Co-Evolutionary Alignment}. We propose the CoEA to address the limitations in LLM-based recommendation exploration. This method integrates long-term group preference modeling and dynamic dual-model collaboration to achieve a balance between novelty exploration and preference alignment.
    \item \textbf{Dual-Stable Interest Exploration}. To solve the problem that the lack of group identity in interest modeling leads to exploration deviating from needs, the DSIE processes long-term and short-term behavioral sequences in parallel and constructs user representations that integrate the stability of group identity and the timeliness of individual interests.
    \item \textbf{Periodic Collaborative Optimization}. To address the flaw that the static optimization of dual models cannot continuously integrate incremental data, the PCO enables the \textit{Relevance LLM} and \textit{Novelty LLM} to form a bidirectional closed-loop and sustainable collaborative cycle.
    \item \textbf{Comprehensive Evaluation}. Extensive online and offline experiments have demonstrated
the effectiveness of our proposed framework for serendipitous
recommendation.
\end{itemize}

\section{Related Work}
\subsection{LLMs for Recommendation}
Current scholarly efforts can be broadly classified into two principal directions. The first strategy capitalizes on LLMs' intrinsic aptitude for text processing to fulfill recommendation tasks \cite{harte2023leveraging, he2023large, hou2024large,  gao2023chat, dai2023uncovering}. It has been observed that LLMs demonstrate a noteworthy ability to make recommendations through straightforward in-context learning. This finding indicates that the models, even during pretraining, can discern item relationships and latch onto fundamental collaborative signals, thus inspiring further research into the integration of LLMs within recommendation systems. The second avenue leverages LLMs for data augmentation \cite{wei2024llmrec, liu2024once} and representation learning \cite{ren2024representation}, aiming to bolster traditional recommendation systems.
In addition to optimizing basic recommendation capabilities, LLMs also provide a new perspective for solving the Exploration-and-Exploitation (EE) balance problem \cite{auer2002finite,coppolillo2024relevance,li2023hierarchical,vargas2011rank,zhao2024breaking,bianchi2025beyond} in recommendation systems. EXPLORE \cite{coppolillo2024relevance} constructs a probabilistic user behavior model and integrates relevance with diversity via a copula function to develop a novel recommendation strategy, aiming to maximize the total amount of knowledge users are exposed to.
Google-v1 \cite{wang2025user} combines hierarchical planning with LLM inference scaling, and separates the goals of novelty and user alignment. Google-v2 \cite{wang2025serendipitous} further introduces multimodal large models on this basis, discovers users' new interests through the chain-of-thought strategy. However, these methods still have limitations: they fail to fully utilize the user group preference information contained in users' long-term interests.

\subsection{User Interest Exploration}
Recommender systems, leveraging observed user-item interactions to infer preferences and deliver personalized content, have achieved significant predominance. Deep Interest Network (DIN) \cite{zhou2018deep} introduced the concept of target attention, enabling the learning of attentive weights for each user behavior with respect to a target item. Subsequently, several works have emerged following the DIN framework. For instance, DIEN \cite{zhou2019deep}, DSIN \cite{feng2019deep}, and BST \cite{chen2019behavior} focus on modeling interest evolution. However, a key challenge lies in the inherent closed-loop nature of existing systems \cite{chen2021values}. Training data is primarily derived from past user-item interactions, limiting the system’s ability to explore truly novel interests. Although recent efforts~\cite{wang2024llms} harness the vast world knowledge of LLMs to mitigate this feedback loop and generate more diverse and serendipitous recommendations, the approach of clustering items based on topic relevance and using LLMs to explore interests of each user is prohibitively expensive for industrial recommendation platforms with hundreds of millions of users.

\section{Preliminary}
\subsection{Problem Formulation}
The core objective of our CoEA is to establish a robust interest exploration mechanism that balances two competing goals: maximizing the novelty of recommended categories relative to users' preferences and ensuring strict alignment with their intrinsic interests.
Formally, we consider user \(u\)'s interaction sequences and their sparse features \(\mathcal{S}_u\) (e.g., age, gender). \(\mathcal{S}_u\) can alleviate the issue of limited interactions for cold-start users.
Formally, we consider a user $u$'s interaction sequences. There are two types of interaction sequences: the long-term click sequence $I_{\text{long}} = \{I^l_1, \ldots, I^l_{N-K}\}$ and the short-term click sequence \(I_{\text{short}} = \{I^s_{N-K+1}, \ldots, I^s_N\}\), which consists of the most recent $K$ interactions.
We aim to learn a mapping function $f: (u, \mathcal{C}_s) \rightarrow \mathcal{C}_n$. Here, $\mathcal{C}_s = \{C_1, \ldots, C_K\}$ represents the set of item categories extracted from the short-term click sequence $I_{\text{short}}$. $\mathcal{C}_n = \{C_{n_1}, \ldots, C_{n_M}\}$ represents the set of novel candidate item categories, and $M$ is the number of categories in $\mathcal{C}_n$. These categories are ones that the user may be interested in but has not interacted with in the short term.

\subsection{Causal Self-Attention}
\label{subsec:causal_attn}
Causal self-attention \cite{yang2021causal} maintains temporal ordering constraints in sequence modeling across $L$ layers by ensuring that position \(i\) cannot attend to any subsequent positions \(j > i\) at each layer. For an input sequence \(\bm{H}^{(l-1)} \in \mathbb{R}^{T \times d}\) at the \(l\)-th layer (where \(1 \leq l \leq L\)), query (\(\bm{Q}\)), key (\(\bm{K}\)), and value (\(\bm{V}\)) matrices are generated through three learnable projections \(\bm{W}_Q, \bm{W}_K, \bm{W}_V \in \mathbb{R}^{d \times d_k}\), with each matrix derived by multiplying the input sequence by its corresponding projection matrix.
Attention weights at each layer are computed under temporal constraints using these three matrices:
\begin{equation}
\bm{A} = \text{softmax}( \frac{\bm{Q}\bm{K}^T}{\sqrt{d_k}} + \bm{M} ) \bm{V},
\label{eq:15}
\end{equation}
where the scaled dot product of query and key matrices is computed by dividing their product by \(\sqrt{d_k}\). A causal mask \(\bm{M}\) — with \(\bm{M}_{ij} = 0\) if \(i \geq j\) and \(\bm{M}_{ij} = -\infty\) if \(i < j\) — is added to enforce autoregressive properties. Softmax is applied to this result, which is then multiplied by the value matrix to generate attention output \(\bm{A}\).
This mask prevents future information leakage, ensuring temporal consistency across \(L\) layers. Each layer’s output is \(\bm{H}^{(l)} = \text{LayerNorm}\left( \bm{A} + \bm{H}^{(l-1)} \right)\). For the final token at position \(T\), its layer-\(l\) representation is \(\bm{H}^{(l)}_T\); after \(L\) layers, \(\bm{H}^{(L)}_T\) encodes the entire sequence’s hierarchical context.
\subsection{RQ-VAE for Group Clustering}
\label{subsec:rq_vae}
Residual-Quantized Variational AutoEncoder (RQ-VAE) \cite{rajput2023recommender} provides hierarchical vector quantization for efficient representation learning. Given user representations \(\bm{u} \in \mathbb{R}^d\) encoded from long-term sequences, we map them to discrete Group Collaborative Semantic IDs (Group CSID) that capture stable group preferences.

The encoder \(\mathcal{E}\) compresses continuous vectors into a latent embedding \(\bm{z} \in \mathbb{R}^{d_z}\) through \(\bm{z} = \mathcal{E}(\bm{u};\theta_E)\). Quantization then occurs through \(K\) residual steps, where each step \(k\) selects the closest codebook vector to the current residual as \(\bm{q}^{(k)} = \underset{\bm{c} \in \mathcal{C}^{(k)}}{\arg\min} \|\bm{r}^{(k)} - \bm{c}\|_2\),
\begin{equation}
\bm{r}^{(k+1)} = \bm{r}^{(k)} - \bm{q}^{(k)},
\end{equation}
where \(\bm{r}^{(0)} = \bm{z}\) initializes the residual.
The Group CSID combines quantization indices into \(CSID(\bm{u}) = (i^{(1)}, i^{(2)}, ..., i^{(K)})\), with \(i^{(k)}\) denoting the codebook index of \(\bm{q}^{(k)}\). Reconstruction uses decoder \(\mathcal{D}\) to map the quantized embedding $\bm{\hat{z}} = \sum_{i = 1}^{K} \bm{q}^{(i)}$ back to the input space as \(\hat{\bm{u}} = \mathcal{D}(\bm{\hat{z}};\theta_D)\).

\begin{figure*}[t]
\centering
\includegraphics[width=0.85\textwidth]{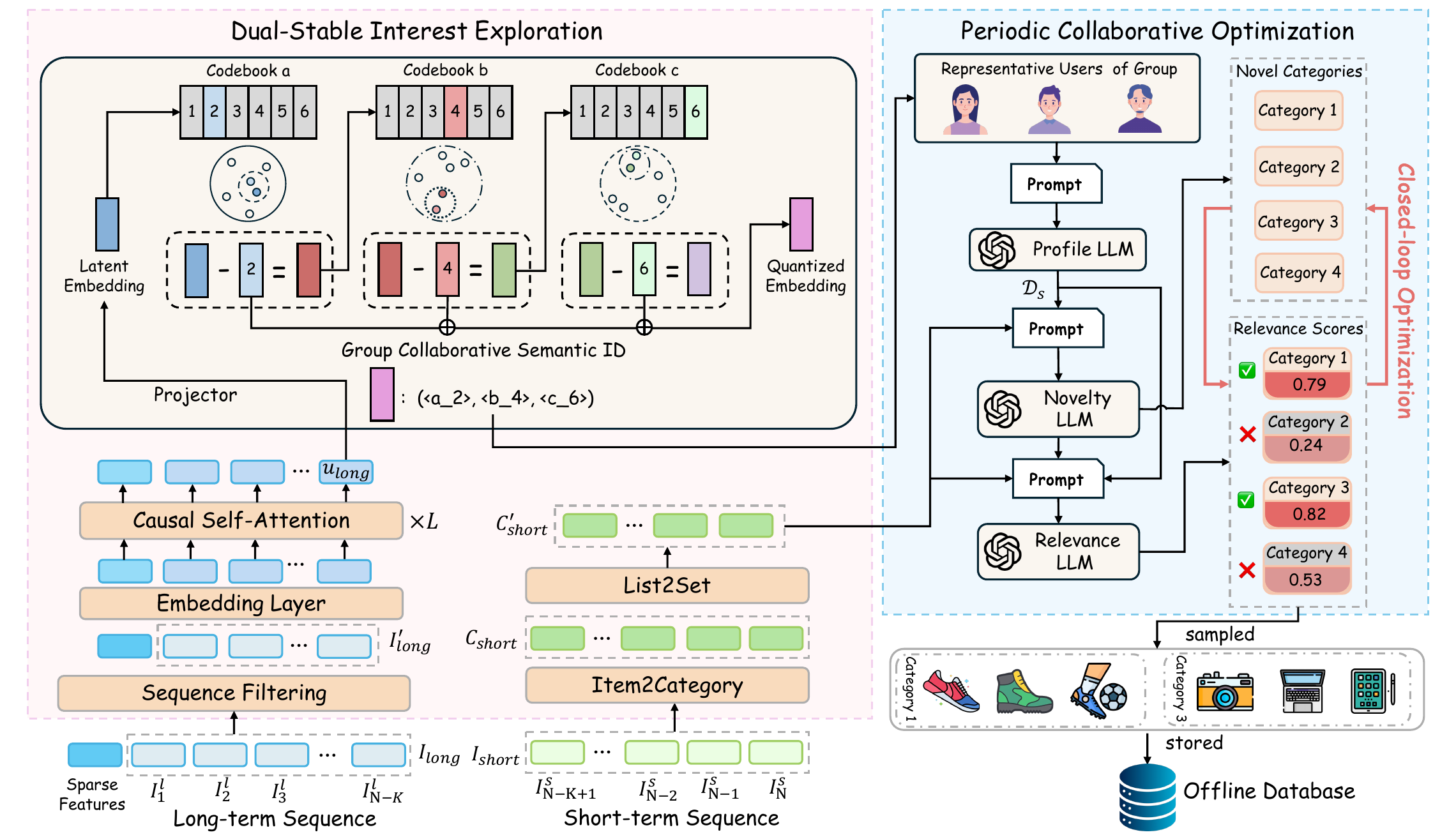} 
% \vspace{-1cm}
\caption{The framework of our CoEA method.}
\label{model}
\end{figure*}

% Long-term and short-term user click sequences are processed by the Dual-Stable Interest Exploration module into user semantic IDs and category info (integrated into a prompt), which is then passed to the  Periodic Collaborative Optimization module—here, potential interest categories generated by \textit{Novelty LLM} are aligned with the user profile via \textit{Relevance LLM}, and items from these categories serve as an additional recall path in the final recommendation model.

\section{Methodology}
In this section, we will describe the proposed CoEA, a method that enables novel interest exploration in recommendation systems while ensuring recommendation quality.

\subsection{Overview}
The overall framework of the CoEA method is shown in Figure \ref{model}. First, the users' long-term and short-term click sequences are input into the DSIE module, which converts the long-term sequences into Group CSID and maps the short-term sequences into fine-grained interest categories. Then, after both are jointly input into the  PCO module, the \textit{Novelty LLM} generates potential novel interest categories by combining the Group CSID with recent interests. These categories will be input into the \textit{Relevance LLM} for real-time preference alignment verification. The alignment results that pass the verification will ultimately be stored in the offline database for subsequent online retrieval.

\subsection{Dual-Stable Interest Exploration}
Existing serendipitous recommendation method \cite{wang2025user} mines novel items that users may be interested in by modeling their short-term behavior sequences. However, such methods fail to fully capture the long-term stable preferences implied by users' group identities from short-term sequences. To address this issue, we propose the Dual-Stable Interest Exploration (DSIE) module
\subsubsection{Sequence Filtering}
\label{sec:seq_filter}
To accurately capture users' core interests and enhance model processing efficiency, we propose a filtering strategy for the long-term sequence \(I_{\text{long}} = \{I^l_1, \ldots, I^l_{N-K}\}\). Given that long-term sequences accumulate abundant historical interaction data, they are well-suited for filtering out occasional behaviors through user click frequencies.
Specifically, we retain items with sufficient user clicks, defined as:
\begin{equation}
I'_{\text{long}} = \left\{ I^l_t \in I_{\text{long}} \mid \text{click}_u(I^l_t) \geq \tau \right\},
\label{eq:9}
\end{equation}
where \(\text{click}_u(i_t)\) represents the cumulative clicks of user \(u\) on item \(i_t\), and \(\tau\) is filtering threshold. The length of \(I'_{\text{long}}\) is denoted as \(T\).

\subsubsection{Long-Term Interest Modeling}
\label{sec:long_term_modeling}
We prepend sparse features \(\mathcal{S}_u\) to the filtered long-term sequence \(I'_{\text{long}}\) and input the combined sequence into causal self-attention (CSA) for joint processing. Each item \(i_t \in I'_{\text{long}}\) is first mapped to a \(d\)-dimensional embedding via:
\begin{equation}
\bm{e_t} = \bm{W}_{emb} \cdot \text{onehot}(i_t),
\label{eq:12}
\end{equation}
where \(\bm{W}_{emb} \in \mathbb{R}^{d \times \mathcal{I} }\) is the trainable embedding matrix and \(\mathcal{I}\) denotes the item vocabulary size. 

Sparse user attribute features \(\mathcal{S}_u\) are embedded into a dense vector \(\bm{e}_{\text{sparse}} \in \mathbb{R}^d\) and prepended to the item embedding sequence \(\bm{E} = [\bm{e_1}, \ldots, \bm{e_T}]\), forming the enhanced input sequence \(\bm{E}'=[\bm{e}_{\text{sparse}}, \bm{e_1}, \ldots, \bm{e_T}]\).
This \(\bm{E}'\) is fed into an \(L\)-layer CSA network, with each layer \(l\) updating the input as:
\begin{equation}
\bm{H}^{(l)} = \text{CSA}(\bm{H}^{(l-1)}),
\label{eq:13'}
\end{equation}
where \(\bm{H}^{(0)} = \bm{E}'\). The aggregated interest representation is captured by \(\bm{u}_{\text{long}} = \bm{H}_{T}^{(L)}\) (the final position output of the last layer).

\subsubsection{Collaborative Semantic IDs Generation}
\label{sec:semantic_id}

The core value of CSID lies in breaking through the limitations of randomness in individual behaviors and exploring temporally and spatially stable interest patterns from a group perspective. Therefore, we cluster user long-term interest representations \(\bm{u}_{\text{long}}\) into collaborative semantic groups using RQ-VAE (defined in Section \ref{subsec:rq_vae}):
\begin{equation}
\text{CSID} = \text{RQ-VAE}(\bm{u}_{\text{long}}).
\label{eq:18}
\end{equation}

For each collaborative semantic group \(\mathcal{G}_s = \{u_i \mid \text{CSID}(u_i) = s\}\), we construct hierarchical profiles: First, $k$ representative users are selected from the nearest neighbors of the cluster centroid \(\bm{\mu}_s\) as follows:
\begin{equation}
\mathcal{R}_s = \underset{u \in \mathcal{G}_s, |\mathcal{R}_s| = k}{\arg\min} \|\bm{u}_{\text{long}} - \bm{\mu}_s\|_2.
\label{eq:representative_users}
\end{equation}

We input the long-term historical behavior sequences  of these $k$ representative users \(\bigcup_{u \in \mathcal{R}_s} I'_{\text{long}}(u)\) into the \textit{Profile LLM} to generate a textual profile \(\mathcal{D}_s\) for this user group $\mathcal{G}_s$ (e.g., "Tech enthusiasts: Prefer flagship mobile phones and foldable screen devices"). The \textit{Profile LLM} is used without fine-tuning, leveraging its native capabilities. The prompt template is provided in Appendix \ref{Appendix:profile}.

\subsubsection{Short-Term Sequence Processing}
\label{short_term}
To directly capture the essential information embedded in users' short-term interest sequences, we adopt a category mapping strategy instead of clustering methods for information consolidation. Specifically, for each item \(I^s_i\) in the short-term sequence \(I_{\text{short}} = \{I^s_{N-K+1}, \ldots, I^s_N\}\), we apply a predefined mapping function \(c(\cdot)\) to transform it into a corresponding category label. This process yields a category list \(\mathcal{C}_{\text{short}} = \big(c(I^s_{N-K+1}), c(I^s_{N-K+2}), \ldots, c(I^s_N)\big)\). Subsequently, we convert this list into a set \(\mathcal{C}'_{\text{short}} = \{c(I^s_i) \mid i = N-K+1,\ldots,N\}\) of length $\mathcal{T}$ to eliminate redundant information and focus on the diversity of categories engaged by the user in recent interactions.

\subsection{Periodic Collaborative Optimization}
Existing methods \cite{wang2025user}, after obtaining feedback from the \textit{Relevance LLM} and completing the alignment of the \textit{Novelty LLM}, only generate a static offline database. This makes them unable to continuously integrate dynamically updated user behavior information and also causes the system to fail to achieve closed-loop optimization based on incremental data. To address this issue, we propose a Periodic Collaborative Optimization (PCO) mechanism: by regularly feeding back user preferences learned by the \textit{Relevance LLM} to the  \textit{Novelty LLM} for incremental alignment, and providing the output of the incrementally fine-tuned  \textit{Novelty LLM} to the \textit{Relevance LLM} for re-scoring, thereby constructing a sustainably evolving dual-model iterative loop.
\subsubsection{Novelty LLM Fine-tuning and Inference}
\label{subsubsec:novelty_llm}
We design distinct prompt construction strategies for the Supervised Fine-Tuning (SFT) \cite{ouyang2022training} and inference stages of the \textit{Novelty LLM}. Specifically, we first process the short-term interaction category sequence $\mathcal{C}'_{\text{short}}$ to obtain $\mathcal{S}_{\text{short}}$, where the last category $\mathcal{S}^{(t)}_{\text{short}}$ has not appeared in the long-term interaction sequence $I_{\text{long}}$. Based on this, we construct prompts for SFT and inference as follows:
\begin{equation}
\mathcal{P}^{(t-2:t-1)}_{\text{ft-nov}} = \mathcal{D}_s \oplus \mathcal{S}^{(t-2:t-1)}_{\text{short}}, \mathcal{P}_{\text{infer-nov}} = \mathcal{D}_s \oplus \mathcal{C}'_{\text{short}},
\label{eq:prompt}
\end{equation}
where the fine-tuning prompt $\mathcal{P}^{(t-2:t-1)}_{\text{ft-nov}}$ combines the user group profile $\mathcal{D}_s$ with the $(t-2)$-th to $(t-1)$-th elements in the short-term category sequence $\mathcal{S}_{\text{short}}$, so as to predict the $t$-th short-term category $\mathcal{S}^{(t)}_{\text{short}}$. The objective function for SFT is:
\begin{equation}
\mathcal{L}_{\text{SFT}} = -\sum_{t=3}^{|\mathcal{S}_{\text{short}}|} \log P_\theta\left(\mathcal{S}^{(t)}_{\text{short}} \mid \mathcal{P}^{(t-2:t-1)}_{\text{ft-nov}}\right).
\label{eq:sft}
\end{equation}

This fine-tuning process learns the fundamental representations of user behavior patterns through autoregressive reconstruction of short-term interest sequences. During inference, we append a novelty trigger instruction to the prompt to guide the model to generate candidate novel categories based on the user profile and historical interactions:
\begin{equation}
\mathcal{C}_{\text{n}} = \text{Novelty\_LLM}(\mathcal{P}_{\text{infer-nov}}),
\end{equation}
where $\mathcal{C}_{\text{n}}$ denotes the set of predicted novel categories.
The fine-tuning configuration and prompt template of \textit{Novelty LLM} are detailed in Appendix \ref{Appendix:novelty}.

\subsubsection{Relevance LLM Fine-tuning and Inference}
\label{subsubsec:relevance_llm}

Drawing inspiration from the RLHF framework \cite{ouyang2022training}, we treat the \textit{Relevance LLM} as a reward model, which is optimized through alignment with human preferences. Differentiated prompts are adopted for the fine-tuning and inference stages of this model, specifically defined as:
\begin{equation}
\mathcal{P}^{(t-2:t-1)}_{\text{ft-rel}} = \mathcal{D}_s \oplus \mathcal{C}'^{(t-2:t-1)}_{\text{short}} \oplus c_{\text{pos/neg}},
\mathcal{P}_{\text{infer-rel}} = \mathcal{D}_s \oplus \mathcal{C}'_{\text{short}} \oplus C_{n_i},
\label{eq:prompt}
\end{equation}
where the novel category \( C_{n_i}\) satisfies \( C_{n_i} \in \mathcal{C}_{\text{n}} \). \( c_{\text{pos}} \) represents the actual clicked category \(\mathcal{C}'^{(t)}_{\text{short}}\) by the user, and \( c_{\text{neg}} \) is a randomly sampled negative sample from the currently exposed but unclicked categories. The corresponding discriminative loss function is:
\begin{equation}
\mathcal{L}_{\text{RM}} = -\mathop{\mathbb{E}}_{(x, c_{\text{pos}}, c_{\text{neg}})} \left[ \log \sigma \big( r_\phi(x, c_{\text{pos}}) - r_\phi(x, c_{\text{neg}}) \big) \right],
\end{equation}
where $x$ denotes the prompt $\mathcal{P}^{(t-2:t-1)}_{\text{ft-rel}}$.
By learning the function \( r_\phi \), the score of positive samples is made higher than that of negative samples, i.e., satisfying \( r_\phi(x, c_{\text{pos}}) > r_\phi(x, c_{\text{neg}}) \). After fine-tuning, the \textit{Relevance LLM} performs inference evaluation on new candidate categories:
\begin{equation}
s_i = \text{Relevance\_LLM}( \mathcal{P}_{\text{infer-rel}} ),
\end{equation}
where the generated alignment score \( s_i \), after normalization, represents the matching degree between the candidate category and user interests. Finally, the target categories are selected through a threshold filtering mechanism:
\begin{equation}
\mathcal{C}_{\text{align}} = \left\{ c_i \in \mathcal{C}_{\text{n}} \mid s_i > \tau_{\text{align}} \right\},
\end{equation}
where \( \tau_{\text{align}} \) is an adjustable confidence threshold. The fine-tuning configuration and prompt template of \textit{Relevance LLM} are detailed in Appendix \ref{Appendix:relevance}.

\subsubsection{Dynamic Closed-loop Optimization}
\label{subsubsec:incremental_tuning}
We perform periodic collaborative updates for the \textit{Novelty LLM} and the \textit{Relevance LLM}. Let the user behavior dataset stored in the system at the beginning of the \( k \)-th cycle be \( \mathcal{D}_k \). Uniformly sample a 1\% subset from it:
\begin{equation}
    \mathcal{D}_k^{\text{sub}} = \left\{ (\mathcal{P}_j, c_{\text{pos},j}, c_{\text{neg},j}) \mid j = 1, \ldots, N \right\}, \ \ N = \lfloor 0.01 \times |\mathcal{D}_k| \rfloor,
\end{equation}
where $\mathcal{P}$ represents the union of $\mathcal{P}_{\text{infer-nov}}$ and the incremental positive categories from the $(k-1)$-th cycle. $c_{\text{pos}}$ and $c_{\text{neg}}$ are the categories with the higher and lower relevance scores predicted by the \textit{Relevance LLM} in the current $k$-th cycle, respectively. Based on the Direct Preference Optimization (DPO) method \cite{rafailov2023direct}, incremental fine-tuning is performed on the \textit{Novelty LLM} with reference to the fixed reference model $\pi_{\text{ref}}$ composed of the parameters of the \textit{Novelty LLM} from the $(k-1)$-th cycle. Its optimization objective is:
\begin{equation}
    \mathcal{L}_{\text{DPO}} = -\mathbb{E}_{\mathcal{D}_k^{\text{sub}}} [ \log \sigma ( \beta ( \log \frac{\pi_\theta(c_{\text{pos}} \mid \mathcal{P})}{\pi_{\text{ref}}(c_{\text{pos}} \mid \mathcal{P})} - \log \frac{\pi_\theta(c_{\text{neg}} \mid \mathcal{P})}{\pi_{\text{ref}}(c_{\text{neg}} \mid \mathcal{P})} ) ) ],
\end{equation}
where the temperature coefficient \( \beta \) controls the strength of the strategy deviating from the reference model.

To avoid catastrophic forgetting \cite{kar2022preventing,korbak2022reinforcement} of the \textit{Novelty LLM} during the incremental fine-tuning process, we introduce the KL divergence to constrain parameter updates:
\begin{equation}
    \mathcal{L}_{\text{TOTAL}} = \mathcal{L}_{\text{DPO}} + \alpha D_{\mathrm{KL}}\left(\pi_\theta(c_{\text{pos}} \mid \mathcal{P})\parallel\pi_{\text{ref}}(c_{\text{pos}} \mid \mathcal{P})\right),
\end{equation}
where \(\alpha\) is the loss coefficient. The incrementally fine-tuned \textit{Novelty LLM} generates novel categories that are more aligned with user preferences. These newly generated categories are fed back to the \textit{Relevance LLM} for scoring; in turn, the evaluation results of the \textit{Relevance LLM} drive further optimization of the \textit{Novelty LLM}. The two thus form a dynamic closed loop, enabling continuous evolution based on incremental data.
\begin{table}[tbp]
  \centering
  \caption{Statistics of datasets.}
  \setlength{\tabcolsep}{4pt} % 调整列间距，可根据需要修改
   \renewcommand{\arraystretch}{0.8}
  \begin{tabular}{lcccc}
    \toprule
    Datasets & \#users & \#items & \#interactions & \#categories \\
    \midrule
    Movielens-1M    & 6,040    & 3,416    & 1,000,209      & 18        \\
    MTRec    & 1,450,381    & 27,953    & 330,722,395      & 594        \\
    \bottomrule
  \end{tabular}
  \label{tab:datasets_statistics}
\end{table}

\begin{table*}[htbp]
  \centering
  \caption{Overall performance on Movielens-1M and MTRec datasets. The last row Improv. indicates the relative improvements of the best performing method (bolded) over the strongest baselines (underlined).}
  % 调整列数为 10 列（Dataset、Metric + 8 个方法）
  % \setlength{\tabcolsep}{3.5pt}
  \renewcommand{\arraystretch}{0.8}
  \begin{tabular}{llcccc|cccc}
    \toprule
    \multirow{2}{*}{\textbf{Type}}& \multirow{2}{*}{\textbf{Method}} & \multicolumn{4}{c}{\textbf{Movielens-1M}} & \multicolumn{4}{c}{\textbf{MTRec}} \\
    \cmidrule(lr){3-6}  
    \cmidrule(lr){7-10}
     & 
    & \textbf{C-H@1} & \textbf{C-N@5} & \textbf{NCP@5} & \textbf{CLTP@5} 
   & \textbf{C-H@10} & \textbf{C-N@10} & \textbf{NCP@10} & \textbf{CLTP@10}     \\  
    \midrule
    \multirow{4}{*}{Exploitation} 
& Two-tower   & \underline{0.8949} & 0.1356 & 0.0394 & 0.0375  & 0.7945 & 0.3741 & 0.0014 & 0.0007  \\
    & GRU4Rec   & 0.8791 & 0.1255 & 0.0619 & 0.0322  & 0.7893 & 0.3759 & 0.0010 & 0.0004  \\
    & SASRec   & 0.8923 & 0.1361 & 0.0590 & 0.0491 & 0.7887 & 0.3681 & 0.0012 & 0.0003  \\
    & Bert4Rec  & 0.8919 & \underline{0.1375} & 0.0349 & 0.0378  & \underline{0.8014} & \underline{0.3855} & 0.0015 & 0.0004  \\

    \midrule
    \multirow{12}{*}{Exploration} 
    & HUCB    & 0.8096 & 0.0966 & 0.0703 & 0.0580   & 0.6904 & 0.2995 & 0.0029 & 0.0008  \\
    & pHUCB  & 0.8173 & 0.1043 & 0.0860 & 0.0599  & 0.7082 & 0.3148 & 0.0026 & 0.0010  \\
    &  NLB  & 0.8045 & 0.0955 & 0.0794 & 0.0508   & 0.7259 & 0.3294 & 0.0037 & 0.0013  \\
    & LLM-KERec  & 0.8848 & 0.1243 & 0.0803 & 0.0680  & 0.7729 & 0.3672 & 0.0045 & 0.0019  \\
    & EXPLORE  & 0.8872 & 0.1282 & 0.1033 & 0.0836  & 0.7852 & 0.3703 & 0.0059 & 0.0027  \\
    & Google-v1   & 0.8854 & 0.1262 & \underline{0.1394} & \underline{0.1204}  & 0.7843 & 0.3690 & \underline{0.0077} & \underline{0.0034}  \\
    & Google-v2 & 0.8376 & 0.0962 & 0.1294 & 0.1184  & 0.7484 & 0.3174  & 0.0072 & 0.0030  \\
   \cmidrule(lr){2-10}  
   & CoEA (w/o Long) & 0.8749 & 0.1158 & 0.1504 & 0.1387  & 0.7835 & 0.3748  & 0.0083 & 0.0042  \\
   & CoEA (w/o Short) & 0.8901 & 0.1305 & 0.1499 & 0.1372  & 0.7983 & 0.3814  & 0.0082 & 0.0040  \\
   & CoEA (w/o R-LLM) & 0.8657 & 0.1124 & 0.1529 & 0.1393  & 0.7803 & 0.3710  & 0.0086 & 0.0046  \\
    & CoEA & \textbf{0.8996} & \textbf{0.1387} & \textbf{0.1535} & \textbf{0.1397} & \textbf{0.8069} & \textbf{0.3887} & \textbf{0.0087} & \textbf{0.0047}  \\
    & Improv. & +0.53\% & +0.87\% & +10.11\% & +16.03\% & +0.69\% & +0.83\% & +12.99\% & +38.24\% \\
    
    \bottomrule
  \end{tabular}
  \label{tab:experiment_results_updated}
\end{table*}
\subsubsection{Offline Storage}
\label{subsubsec:offline_storage}
To reduce the computational pressure on the online serving system, we adopt an offline storage strategy for learned novel categories: all aligned categories \(\mathcal{C}_{\text{align}}\) are persistently stored in a distributed key-value database, decoupling the exploration process from online inference. Specifically, we construct composite keys by concatenating the Collaborative Semantic ID (CSID) with the hashed representation of 
\(\mathcal{C}_{\text{short}}\); the corresponding values are used to store the novel category list \(\mathcal{C}_{\text{align}}\). This design enables efficient online retrieval during the serving process through simple database queries alone.

\section{Experiments}

In this section, we present empirical results to demonstrate the effectiveness of our proposed CoEA. These experiments are designed to answer the following research questions:
\begin{itemize}
    \item \textbf{RQ1} How does CoEA perform compared with state-of-the-art recommendation models?
    \item \textbf{RQ2} What is the effect of Dual-Stable Interest Exploration and  Periodic Collaborative Optimization in our proposed CoEA?
    \item \textbf{RQ3} How does CoEA perform in the real-world online recommendations with practical metrics?
    \item \textbf{RQ4} How do hyper-parameters in CoEA impact recommendation performance?
\end{itemize}
\subsection{Experimental Settings}
\subsubsection{Datasets}

To validate the effectiveness of our method, we conduct evaluations on two real-world datasets: Movielens-1M \footnote{https://grouplens.org/datasets/movielens/1m/} and MTRec. Movielens-1M is a movie rating dataset with 18 movie categories, and its data processing refers to \cite{sun2019bert4rec}. The MTRec dataset is collected from real recommendation scenarios on the homepage of Meituan App (Meituan is one of the largest takeaway platforms in China). It contains 25 days of user click behavior data from April 2 to 26, 2025, in a certain city, covering 594 item categories. This dataset is divided by time into a training set (first 20 days), a validation set (21th-23th day), and a test set (last two day). Detailed
statistics of the datasets are shown in Table \ref{tab:datasets_statistics}.

\subsubsection{Metrics} We employ different evaluation metrics across offline and online experiments. For offline evaluation, category-level metrics are prioritized: recommendation quality is quantified by Category Hit Rate (C-H@K) and Category Normalized Discounted Cumulative Gain (C-N@K) to capture user preference alignment, while novelty is measured via self-defined metrics including Novel Category Proportion (NCP@K) and Category Long-Tail Proportion (CLTP@K). In online settings, business metrics are adopted: Gross Transaction Value (GTV) reflects quality impact, and 7-Day Novel Item Exposure Proportion (7D-NIEP) assesses novelty in real-world deployments. Comprehensive metric definitions and computational details are documented in Appendix \ref{Appendix:metrics111}.
\begin{figure}[t]
\centering
\subfigure[Movielens-1M]{
    \includegraphics[width=0.45\columnwidth]{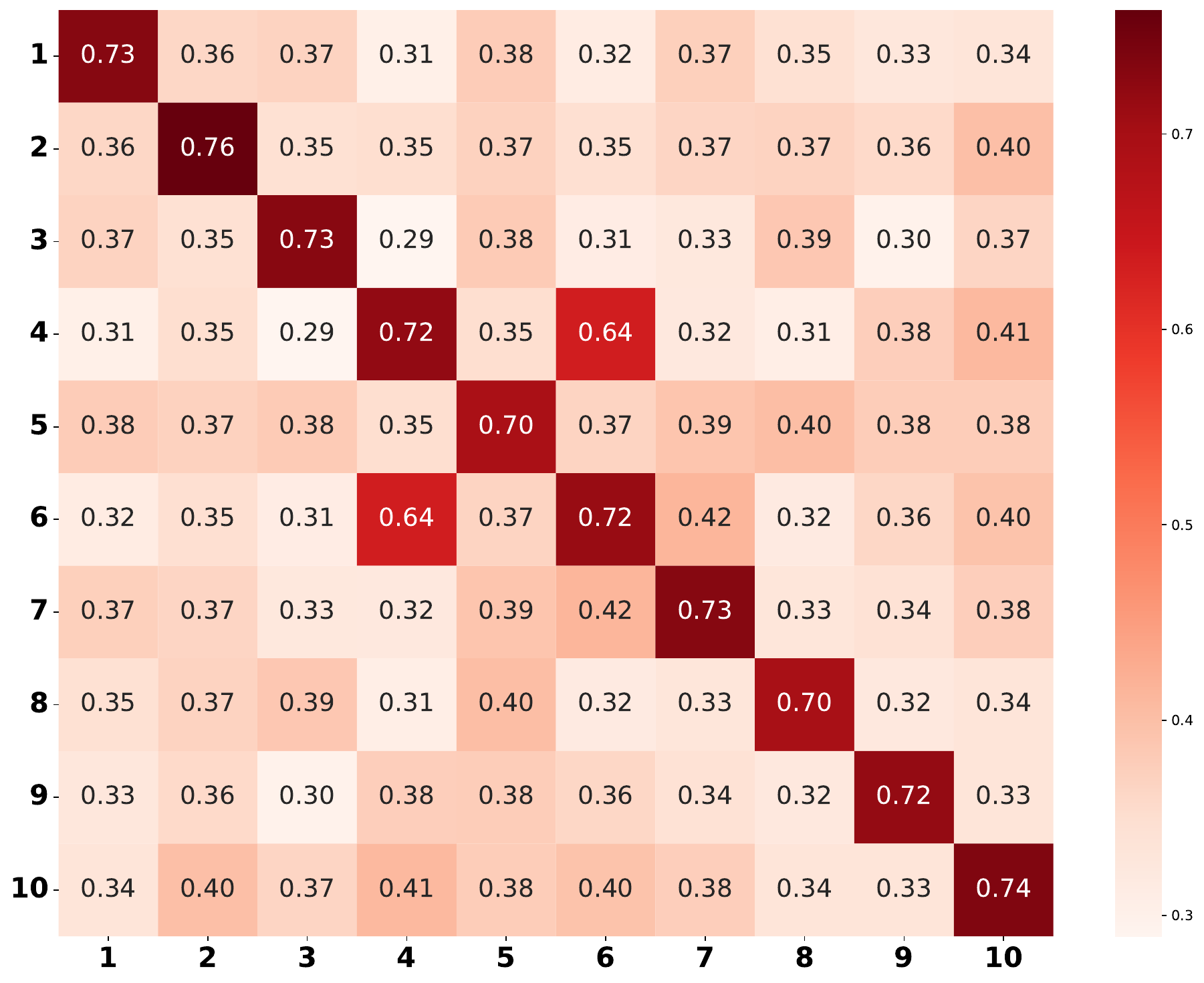}
    \label{heatmap1}
}
% \hfill
\subfigure[MTRec]{
    \includegraphics[width=0.45\columnwidth]{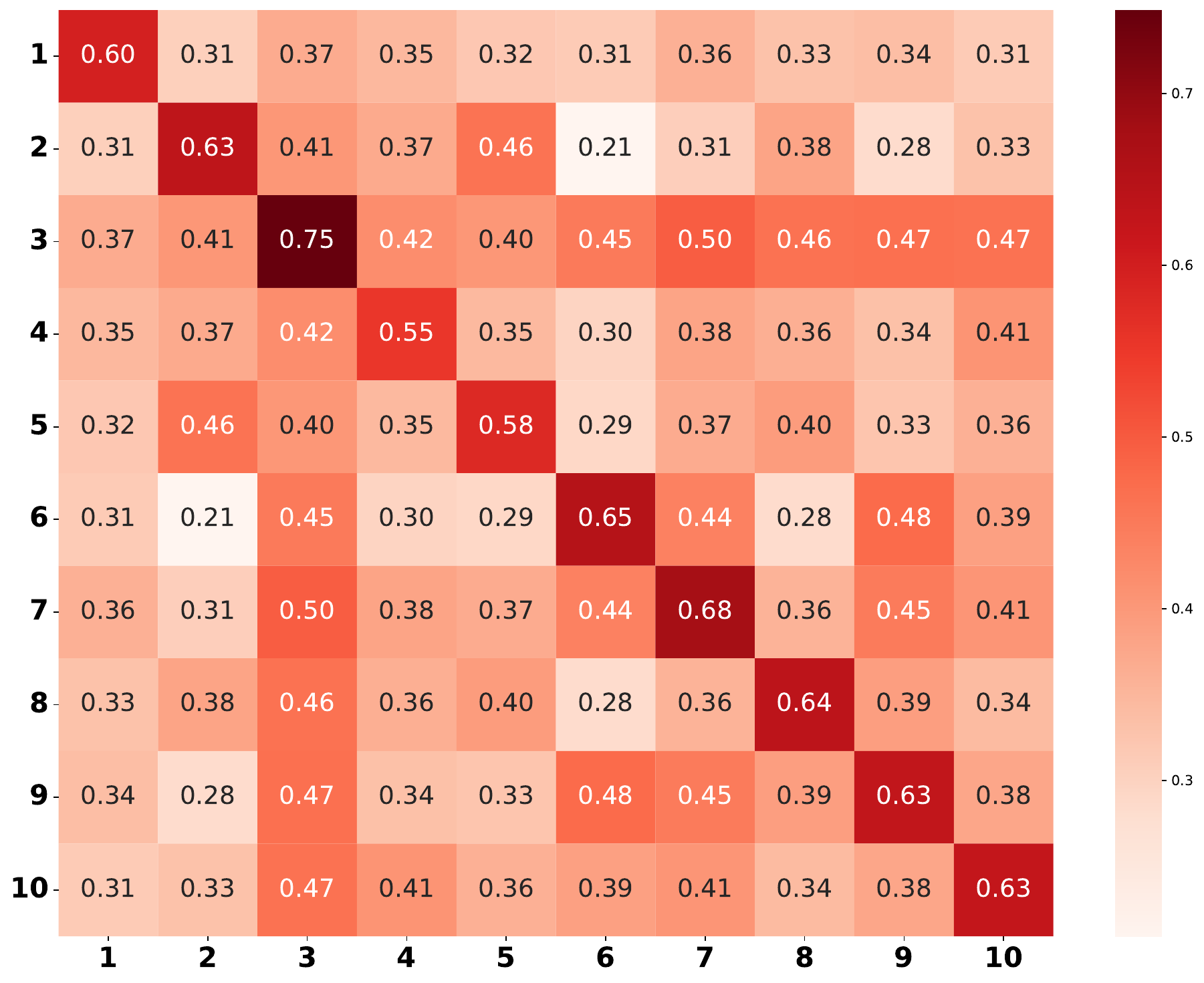}
    \label{heatmap2}
}
\caption{Heatmaps of inter-group and intra-group user similarity for the Movielens-1M and MTRec datasets.}
\label{heatmap}
\end{figure}
\subsection{Overall Performance Comparison (RQ1)}
We compare our CoEA with 4 exploitation-based baseline models (i.e., Two-tower \cite{yang2020mixed}, GRU4Rec \cite{hidasi2015session}, SASRec \cite{kang2018self}, and Bert4Rec \cite{sun2019bert4rec}) and 7 exploration-based baseline models (i.e., HUCB/pHUCB \cite{song2022show}, NLB \cite{su2024long}, LLM-KERec \cite{zhao2024breaking}, EXPLORE \cite{coppolillo2024relevance}, Google-v1 \cite{wang2025user}, and Google-v2 \cite{wang2025serendipitous}). Since the two datasets do not contain multimodal information, we use a text LLM instead of a multimodal LLM when conducting experiments on the Google-v2 model.
Details of the baseline models are available in Appendix \ref{Appendix:baselines}, and implementation details of CoEA are in Appendix \ref{appendix:detail}. The experimental results are the average value obtained after running five times. Table \ref{tab:experiment_results_updated} presents the overall performance on the two datasets, with the result analysis as follows:
\begin{itemize}
    \item The CoEA model achieves significant performance improvements across multiple benchmark datasets. Compared to the best-performing baseline, it demonstrates an average gain of 0.73\% on category-level quality metrics (C-H@K and C-N@K) and a remarkable 19.34\% improvement on category-level novelty metrics (NCP@K and CLTP@K). These results confirm that CoEA effectively resolves the core challenge of balancing quality and diversity in traditional recommendation systems, particularly excelling in long-tail category discovery with a 27.14\% boost in CLTP metrics.
    \item Exploration-based baselines exhibit a clear performance trade-off when compared to exploitation-based approaches: while achieving substantial gains in novelty metrics (NCP@K and CLTP@K), they generally underperform in quality metrics (C-H@K and C-N@K).
    \item Google-v1 shares architectural similarities with CoEA, as both employ a dual-model framework combining \textit{Novelty LLM} and \textit{Relevance LLM}, whereas Google-v2 utilizes a single novelty-focused multimodal LLM. Leveraging the human-behavior alignment capability of its \textit{Relevance LLM}, Google-v1 significantly outperforms Google-v2 on quality metrics (C-H@K and C-N@K). Notably, CoEA surpasses Google-v1 across all metrics, demonstrating its ability to maintain novelty advantages while further enhancing recommendation relevance through multi-cycle incremental fine-tuning.
\end{itemize}

\subsection{Ablation Study (RQ2)}
Since the Dual-Stable Interest Exploration module and the  Periodic Collaborative Optimization mechanism are the core of CoEA, we validate the effectiveness of these core components through ablation experiments.
\subsubsection{Impact of Dual-Stable Interest Exploration}
For the validation experiment of the Dual-Stable Interest Exploration (DSIE) module, we design ablation tests and construct two variant models: CoEA (w/o Long), which removes the long-term sequence, and CoEA (w/o Short), which removes the short-term sequence.  Experimental results (as shown in Table \ref{tab:experiment_results_updated}) show that CoEA (w/o Long) significantly declines in core quality metrics (C-H@K and C-N@K) with an average drop of 5.9\%, while its novelty metrics (NCP and CLTP) remain stable. In contrast, CoEA (w/o Short) only has slight fluctuations in novelty metrics and no obvious attenuation in core performance metrics. This confirms the key role of DSIE in capturing users' stable preferences.

\begin{table}[tbp]
  \centering
  \caption{The cosine similarity between the category representation and the user representations.}
  \label{tab:random}
    \setlength{\tabcolsep}{10.5pt}
      \renewcommand{\arraystretch}{0.8}
  \begin{tabular}{@{}lcc@{}}
    \toprule
 & Movielens-1M & MTRec \\
    \midrule
     Random &  0.252 & 0.226 \\
     CoEA-Exchange &  0.483 & 0.462 \\
    CoEA & 0.561 &  0.502 \\
    \bottomrule
  \end{tabular}
\end{table}
To further validate the effectiveness of each component in the DSIE module, we conduct a two-phase analysis. In the user clustering validation phase, we employ the Movielens-1M (clustered into 50 groups) and MTRec (clustered into 100 groups) datasets. Figure \ref{heatmap} presents the heatmap of user groups generated by RQ-VAE. For each dataset, we randomly select 10 user groups and sample two subsets of 30 user representations per group. After averaging the representations within each subset, we obtain two representative vectors per group and compute their cosine similarity. The results demonstrate that the average intra-group cosine similarity is significantly higher than inter-group similarity, confirming that users with consistent interests are accurately aggregated. Notably, Movielens achieves higher intra-group similarity than MTRec, attributable to its fewer user groups yielding better clustering performance.

In the model validation stage we design comparative experiments to validate the effectiveness of long-term interest sequence modeling with results shown in Table \ref{tab:random}. Specifically we use \textit{Novelty LLM} to generate 100 item representations for each novel category and obtain category representation vectors via average pooling; we then compute the cosine similarity between these category representations and user long-term interest representations output by Causal Self-attention (CSA). For comprehensive model evaluation we set two control experiments: (1) a Random baseline which calculates similarity between category representations and the average of 10 randomly selected irrelevant user representations; and (2) a CoEA-Exchange baseline which feeds short-term interest sequences (instead of long-term ones) into CSA to derive user representations. Experimental results on both Movielens-1M and MTRec datasets show our method (average similarity 0.532) significantly outperforms the Random baseline (average similarity 0.239) and the CoEA-Exchange baseline (average similarity 0.473) validating that long-term interest sequence modeling captures more stable user preference information enabling the LLMs to more accurately recommend novel categories aligned with user interests.

\begin{figure}[t]
\centering
\subfigure[CoEA (w/o KL)]{
    \includegraphics[width=0.47\columnwidth]{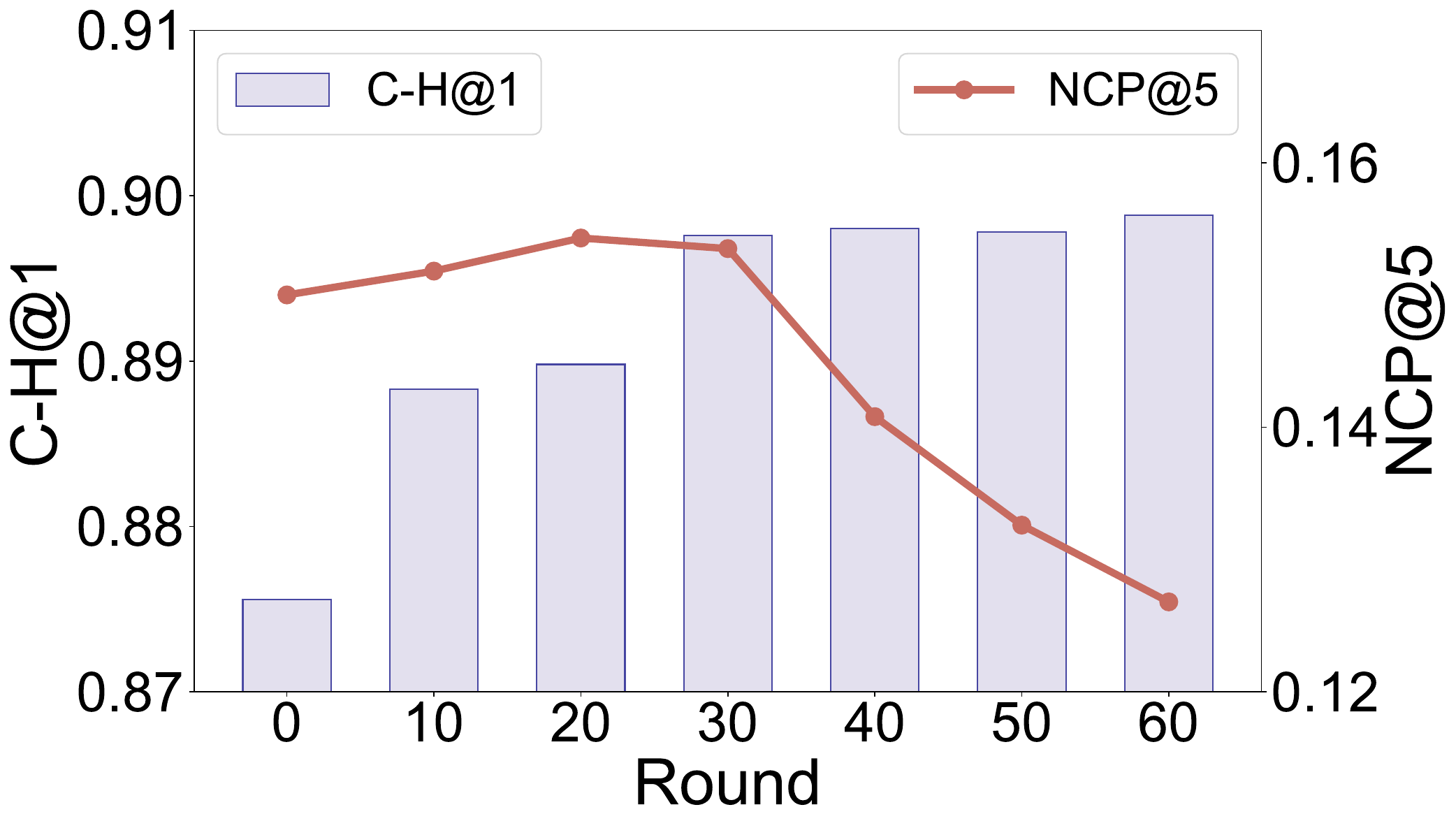}
    \label{dddp}
}
\subfigure[CoEA]{
    \includegraphics[width=0.47\columnwidth]{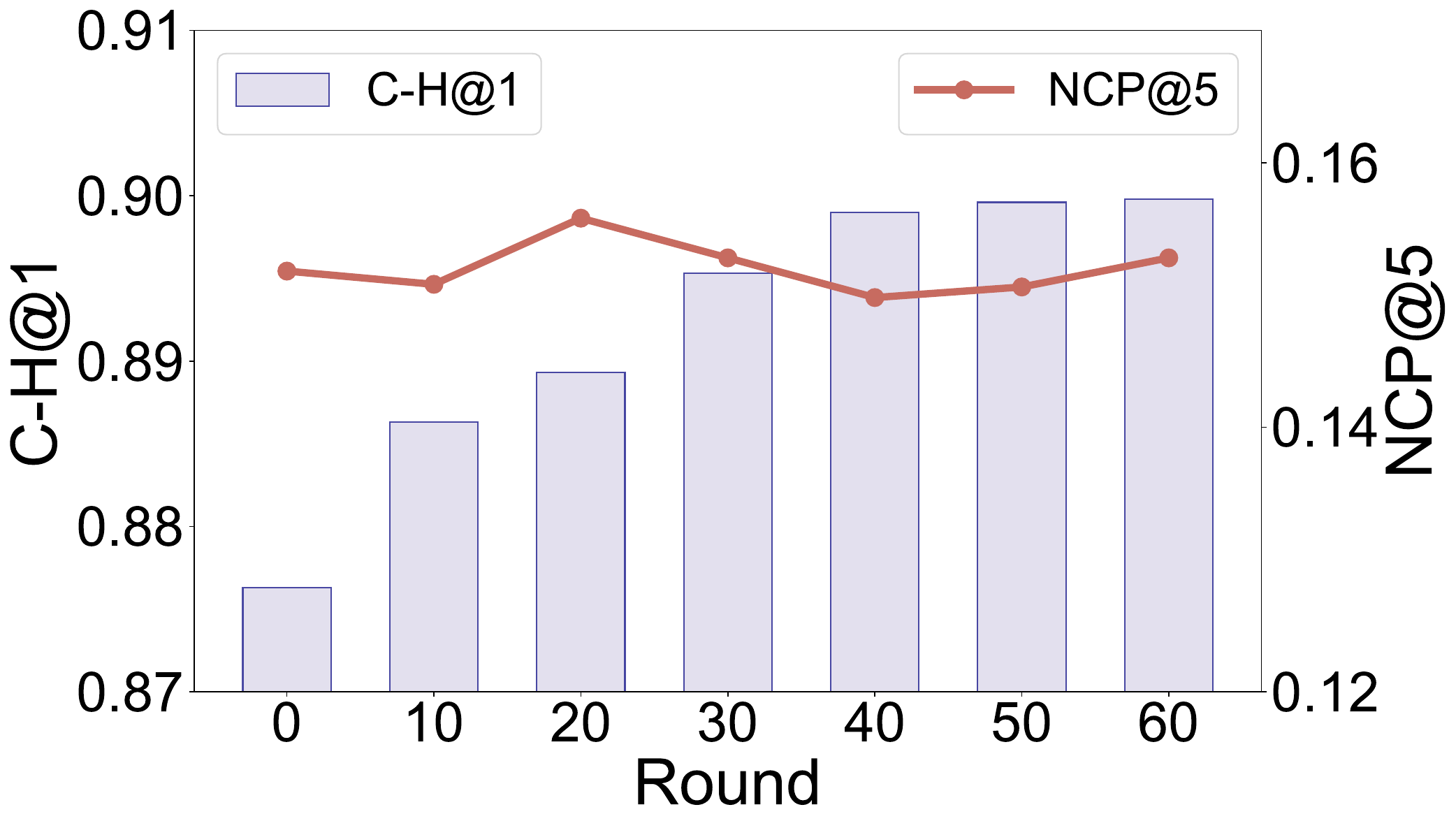}
    \label{second-figure}
}
\caption{Quality and Novelty Metrics Across Fine-tuning Rounds: CoEA vs. CoEA (w/o KL) in Movielens-1M dataset.}
\label{fig:zhuzhuang}
\end{figure}
\subsubsection{Impact of  Periodic Collaborative Optimization}
To validate the effectiveness of the  Periodic Collaborative Optimization (PCO) mechanism, we conduct an ablation study by removing the \textit{Relevance LLM} to construct a variant model CoEA (w/o R-LLM). As shown in Table \ref{tab:experiment_results_updated}, while the novelty metrics remain largely stable, the recommendation quality significantly deteriorates. This demonstrates that relying solely on the \textit{Novelty LLM} leads to quality degradation, whereas the PCO mechanism effectively maintains recommendation relevance through alignment, ensuring suggested items better match user preferences.

We further validate PCO's incremental optimization through multi-round fine-tuning experiments on the Movielens-1M dataset. Comparing the original CoEA and its variant CoEA (w/o KL), i.e., without the KL divergence loss, we conduct 0-60 rounds of incremental fine-tuning (5 batches per round, batch size=1024). Each round feeds back 1\% stored data to \textit{Novelty LLM} while synchronously optimizing \textit{Relevance LLM}. Figure \ref{fig:zhuzhuang} shows that while both models improve quality metrics and maintain novelty stability initially, beyond 30 rounds the KL-ablated variant exhibits stagnated quality and significant novelty degradation, whereas CoEA preserves stability. This confirms the KL divergence loss prevents \textit{Novelty LLM} from strong feedback loops that compromise novelty.

\begin{figure}[t]
\centering
\includegraphics[width=\linewidth]{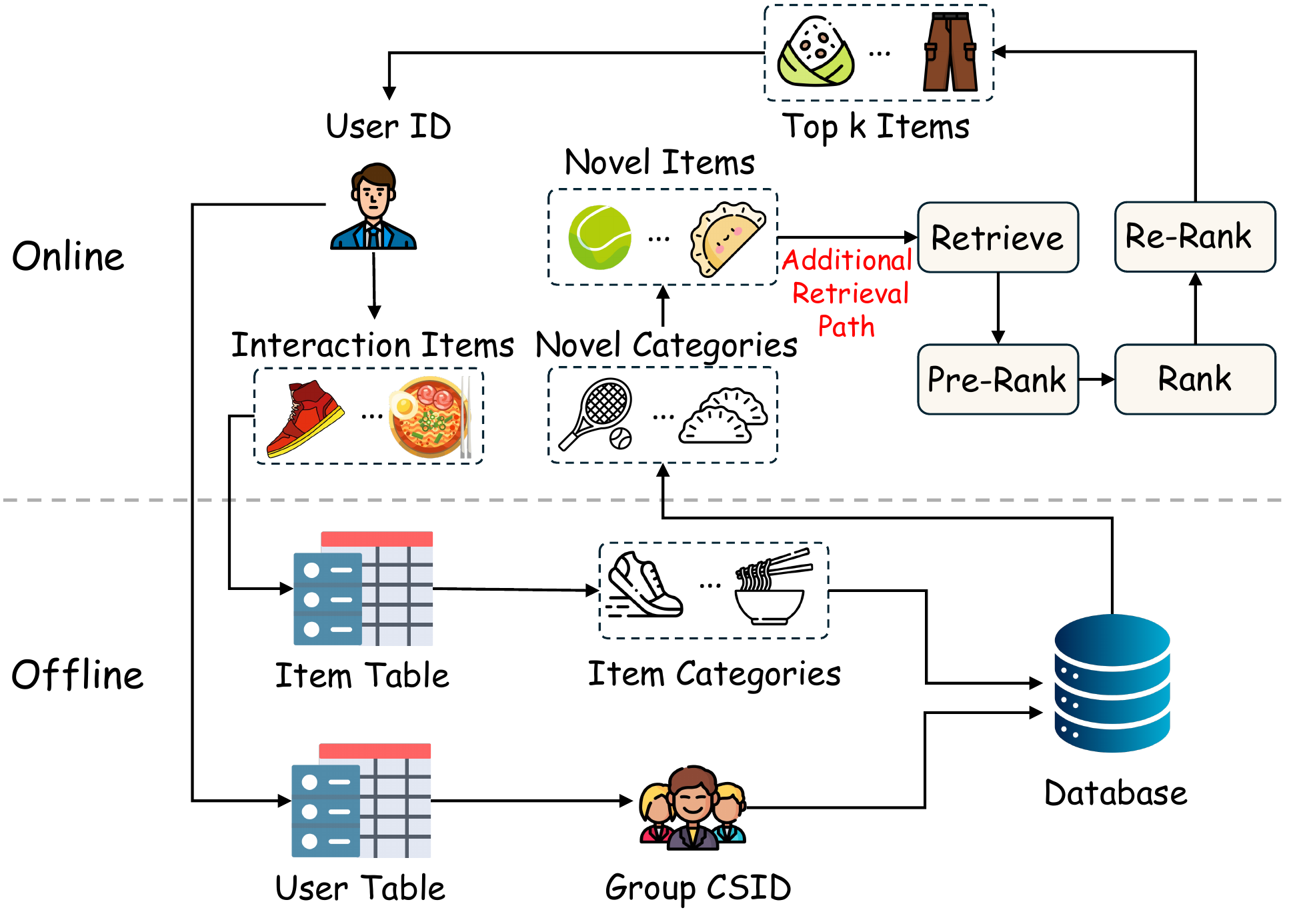} 
% \vspace{-cm}
\caption{Deployment architecture of CoEA in online
recommendation system.}
\label{online}
\end{figure}

\subsection{Online A/B test (RQ3)}
CoEA has been successfully deployed in Meituan App's homepage recommendation system. As shown in Figure \ref{online}, the system processes $T-1$ historical behavior logs offline via Spark while capturing real-time behavior streams through Flink. Upon user request, the feature engine retrieves group CSIDs from offline User Table and item categories from Item Table, queries the Database for novel categories, and generates an additional retrieval path after filtering. Cold-start users are mapped to default group CSID with learning data updated daily.
A 20-day A/B test demonstrates CoEA's significant improvements: +1.203\% GTV and +2.364\% 7D-NIEP. This validates that CoEA effectively models the relationship between users' long-term interests and novel categories, enabling it to discover latent interests while maintaining recommendation relevance, thereby achieving synergistic optimization of both business metrics and novelty.

\subsection{Hyper-Parameter Studies (RQ4)}
\begin{figure}[t]
\centering
\subfigure[Sequence filtering threshold $\tau$]{
    \includegraphics[width=0.22\textwidth]{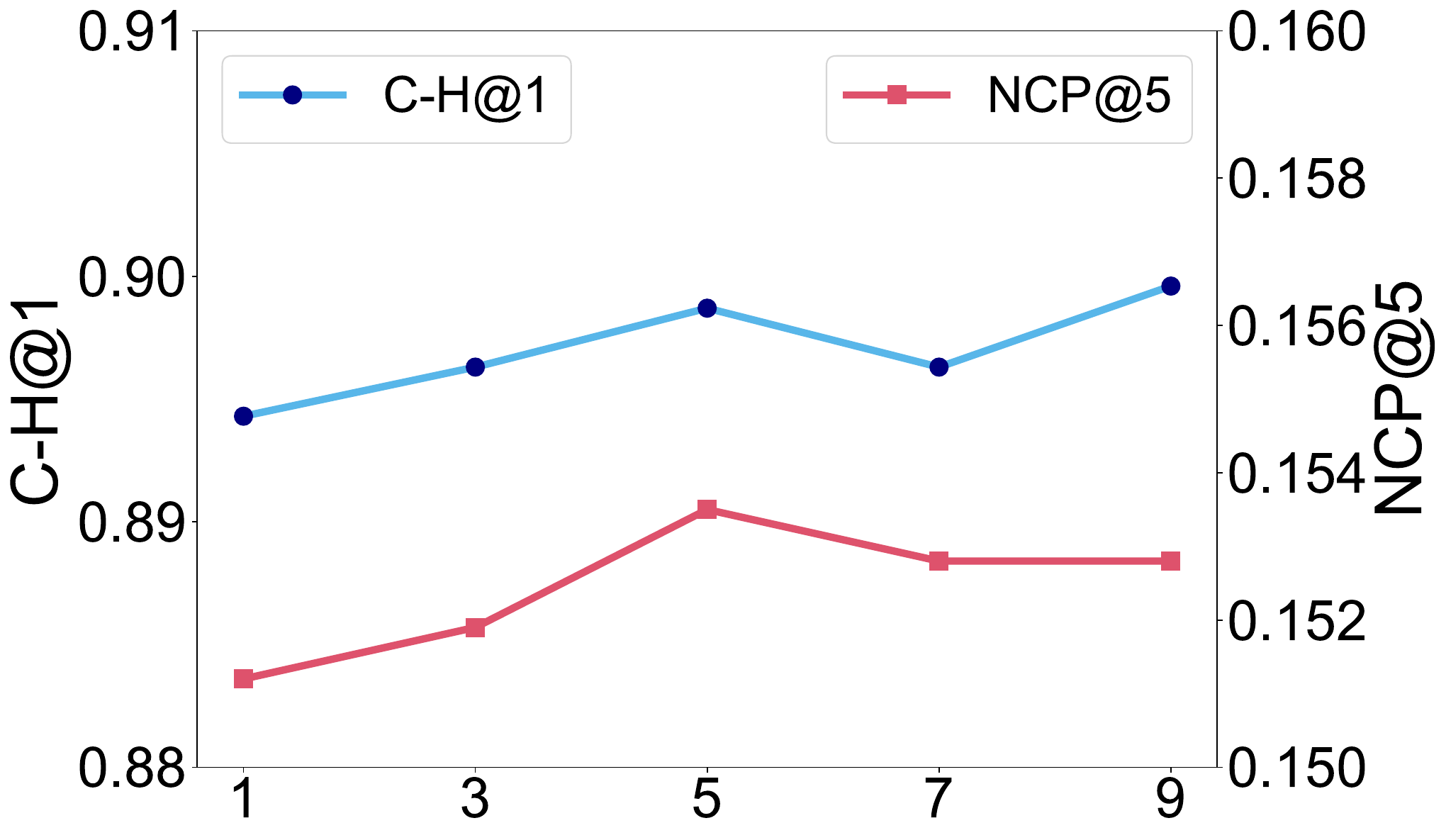}
    \label{ddd}
}
% \hfill
\subfigure[Loss coefficient $\alpha$]{
    \includegraphics[width=0.22\textwidth]{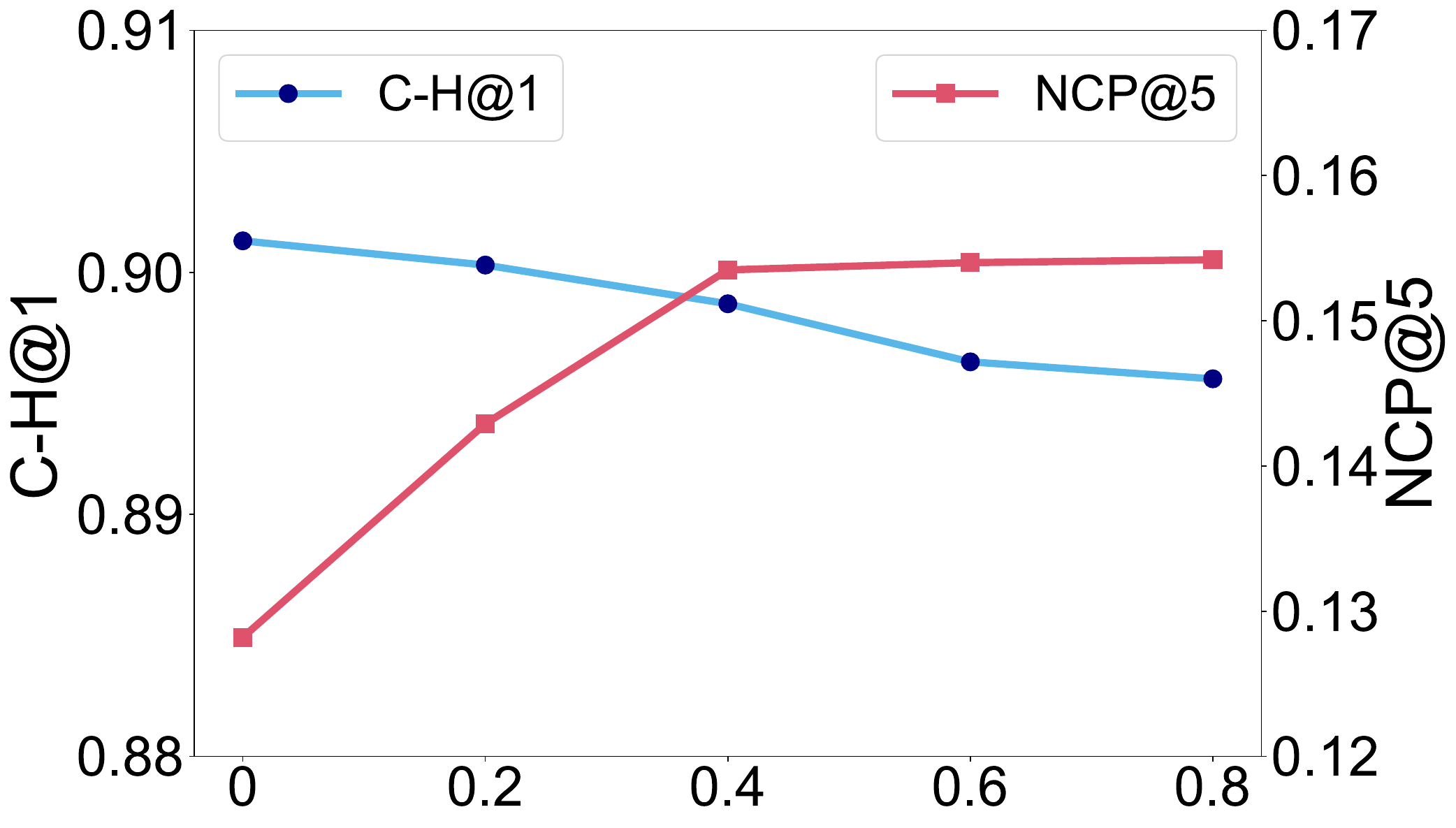}
    \label{second-figure}
}
\subfigure[Number of residual quantizer layers $l$]{
    \includegraphics[width=0.22\textwidth]{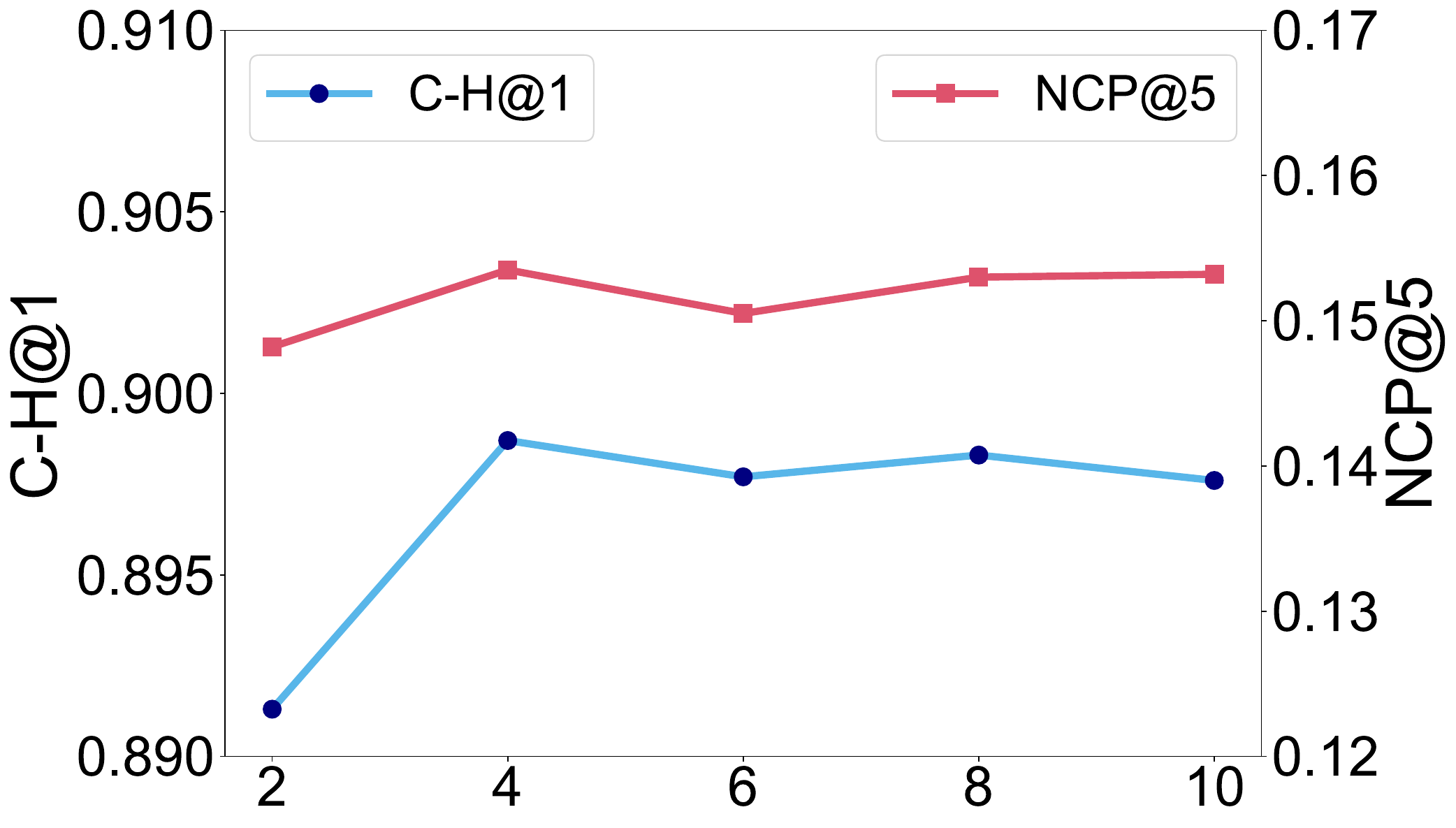}
    \label{second-figure}
}
\subfigure[Number of user groups $M_N$]{
    \includegraphics[width=0.22\textwidth]{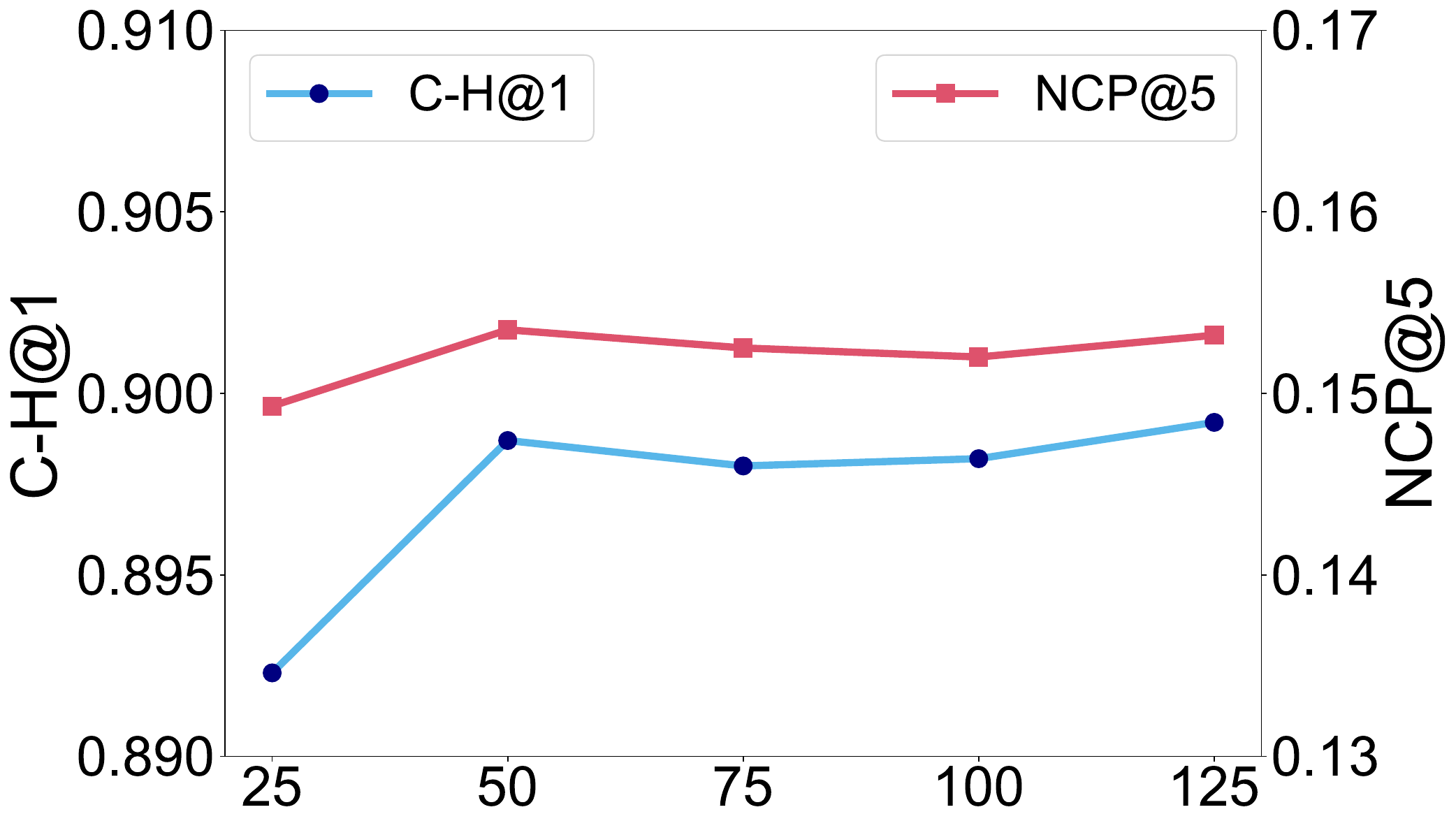}
    \label{second-figure}
}
\caption{Performance of the four parameters with different values on the Movielens-1M dataset.}
\label{fig:para}
\end{figure}

In this section, we conduct experiments on the Movielens-1M dataset to explore the impact of the four parameters, and the experimental results are shown in Figure \ref{fig:para}.

    \textbf{Impact of sequence filtering threshold}. When $\tau$ is in the range of $[5, 9]$, the metrics C-H@1 and NCP@5 show fluctuations, where C-H@1 remains stable above 0.896 and NCP@5 remains stable above 0.152. When $\tau$ is in the range of $[1, 5]$, both C-H@1 and NCP@5 show an upward trend as $\tau$ increases, which indicates that noise interference will lead to a decrease in C-H@1 and C-N@5.  

    \textbf{Impact of loss coefficient}. The model achieves the best stability when $\alpha=0.4$. During 30 rounds of incremental learning: if $\alpha \leq 0.2$, catastrophic forgetting will cause a sharp drop in NCP@5; if $\alpha \geq 0.6$, parameter rigidity will lead to a decrease in C-H@1. The setting of $\alpha=0.4$ realizes the synchronous optimization of quality metrics and novelty metrics.

    \textbf{Impact of number of residual quantizer layers}. When $l$ is in the range of [4,10], the metrics C-H@1 and NCP@5 show fluctuations with no significant changes; when $l=2$, both metrics decrease significantly, indicating that the discrimination of user groups is poor at this time; while when $l$ is greater than 4, the improvement of metrics is not obvious. Therefore, the setting of $l=4$ balances semantic granularity and computational efficiency.

    \textbf{Impact of number of user groups}. When $M_N$ is in the range of [50,125], the metrics C-H@1 and NCP@5 show fluctuations. Under extreme configurations: when $M_N=25$, overly coarse group division will lead to a decrease in both metrics; when $M_N>100$, although the metrics do not decrease, the excessive number of groups will increase online retrieval latency. Therefore, the optimal value is $M_N=50$.
\section{Conclusion and Future Work}
Our proposed CoEA alleviates the Exploration-and-Exploitation dilemma in serendipitous recommendation through two components: the first is DSIE, which can simultaneously capture the stable commonalities at the group level and the dynamic timeliness at the individual level; the second is PCO, which can dynamically optimize the generation of novel content based on real-time relevance signals. Experiments on industrial-scale datasets demonstrate the superior performance of CoEA: in offline evaluation, its recommendation quality and novelty discovery ability are significantly better than those of baseline models; the online deployment in Meituan App further verifies its practical value, successfully achieving a GTV increase of +1.203\% and a 7D-NIEP growth of +2.364\%. For future work, we will focus on optimizing the real-time performance of model updates for cold-start users.
% \newpage

% \begin{acks}
% \end{acks}

\bibliographystyle{ACM-Reference-Format}
\bibliography{sample-base}

\newpage
\appendix

\begin{figure*}[t]
\centering
\subfigure[Profile LLM]{
    \includegraphics[width=0.32\textwidth]{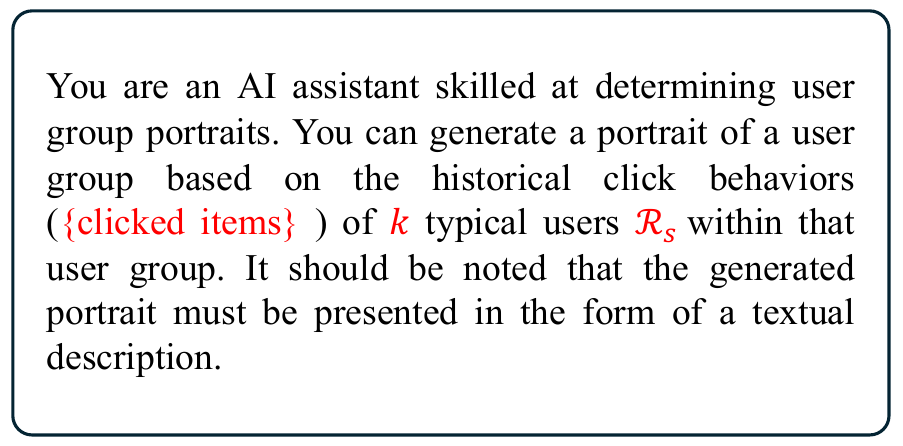}
    \label{prompt_11}
}
% \hfill
\subfigure[Novelty LLM]{
    \includegraphics[width=0.32\textwidth]{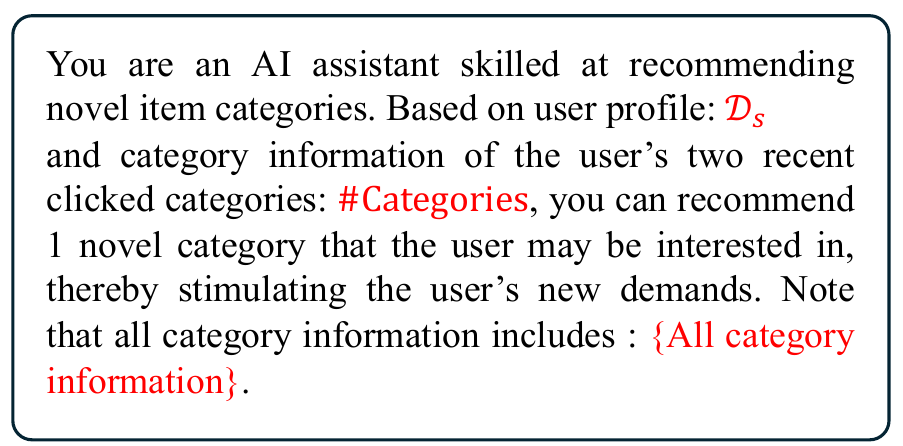}
    \label{prompt_22}
}
\subfigure[Relevance LLM]{
    \includegraphics[width=0.32\textwidth]{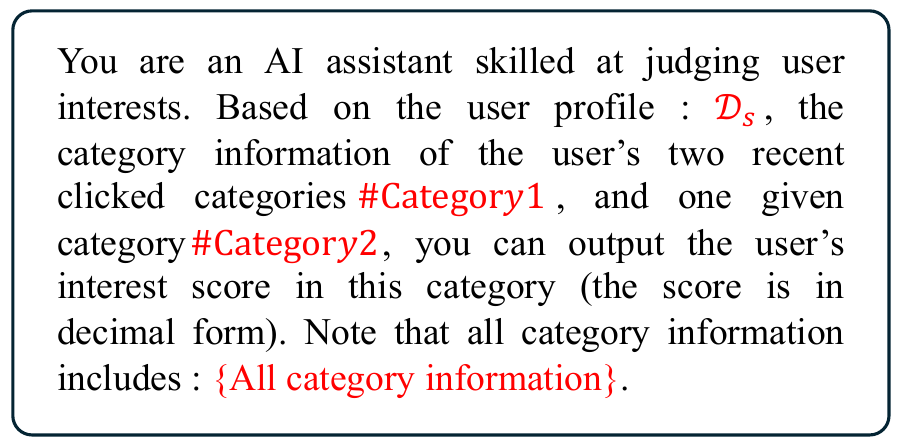}
    \label{prompt_33}
}
\caption{The prompt templates for the three different LLMs.}
\label{prompts}
\end{figure*}
\section{Appendix}
\subsection{Relevant Details of the Profile LLM}
\label{Appendix:profile}
For \textit{Profile LLM}, no fine-tuning is performed, and its native capabilities are directly utilized with the prompt template as shown in Figure \ref{prompt_11}.

\subsection{Relevant Details of the Novelty LLM}
\label{Appendix:novelty}
For \textit{Novelty LLM}, it is trained through supervised fine-tuning based on the Qwen2.5-7B-Instruct-1M model. Its prompt template is shown in Figure \ref{prompt_22}. Its fine-tuning prompt and inference prompt are \(\mathcal{P}^{(t-2:t-1)}_{\text{ft-nov}}\) and \(\mathcal{P}_{\text{infer-nov}}\) respectively, and they share the same prompt template, with differences only in the specific content to be filled during inference: for fine-tuning, the content corresponding to \#Categories in the template is \(\mathcal{S}^{(t-2:t-1)}_{\text{short}}\) from \(\mathcal{P}^{(t-2:t-1)}_{\text{ft-nov}}\); for inference, the content corresponding to \#Categories in the template is \(\mathcal{C}'_{\text{short}}\) from \(\mathcal{P}_{\text{infer-nov}}\). If their lengths are different, the last two categories of \(\mathcal{C}'_{\text{short}}\) are selected to ensure consistent length.

Specifically, the LoRA method (rank = 8, targeting all layers) is adopted for efficient fine-tuning, combined with DeepSpeed to achieve automated training optimization; mixed-precision training enables FP16 (including dynamic loss scaling with an initial scaling exponent of 16, a scaling window of 1000 steps, and a minimum scaling value of 1) and BF16 (both set to "auto"); ZeRO optimization stage 2 improves memory utilization through configurations such as full-partition collection and distributed reduction (with a communication bucket size of 500MB and overlapping communication and computation). The training is conducted with a per-device batch size of 2, gradient accumulation of 2, and a learning rate of 1.0e-4. 

In the incremental fine-tuning of closed-loop optimization, we perform LoRA fine-tuning on \textit{Novelty LLM} using the DPO (Direct Preference Optimization) method, with specific configurations as follows: LoRA rank is 8, covering all layers, the preference value is set to 0.1, and the sigmoid loss function is adopted. Regarding training parameters, the single-device batch size is 2, the number of gradient accumulations is 8, and the learning rate is 5.0e-6.

\subsection{Relevant Details of the Relevance LLM}
\label{Appendix:relevance}
For \textit{Revelance LLM}, it is also based on the Qwen2.5-7B-Instruct-1M model, with fine-tuning configurations consistent with those of \textit{Novelty LLM}.
Its prompt template is shown in Figure \ref{prompt_33}. The fine-tuning prompt \(\mathcal{P}^{(t-2:t-1)}_{\text{ft-rel}}\) and inference prompt \(\mathcal{P}_{\text{infer-rel}}\) share the same template, with differences only in the specific content to be filled during inference: during fine-tuning, \#Category1 in the template corresponds to \(\mathcal{C}'^{(t-2:t-1)}_{\text{short}}\) from \(\mathcal{P}^{(t-2:t-1)}_{\text{ft-rel}}\), and \#Category2 corresponds to \(c_{\text{pos}}\) or \(c_{\text{neg}}\) from it; during inference, \#Category1 in the template corresponds to \(\mathcal{C}'_{\text{short}}\) from \(\mathcal{P}_{\text{infer-rel}}\), and \#Category2 corresponds to \(C_{n_i}\) from it. If \(\mathcal{C}'^{(t-2:t-1)}_{\text{short}}\) and \(\mathcal{C}'_{\text{short}}\) have different lengths, the last two categories of \(\mathcal{C}'_{\text{short}}\) should be selected to ensure consistent length.

% \begin{figure}
% \centering
% % \hfill
% \subfigure[Fine-tune prompt template]{
%     \includegraphics[width=0.8\linewidth]{samples/fig/prompt3_1.pdf}
%     \label{prompt_23}
% }
% \subfigure[Inference prompt template]{
%     \includegraphics[width=0.8\linewidth]{samples/fig/prompt3_2.pdf}
%     \label{prompt_3332}
% }
% \caption{The fine-tuning and inference prompt templates of Relevance LLM.}
% \label{prompt_3}
% \end{figure}

\subsection{Evaluation Metrics}
\label{Appendix:metrics111}
\subsubsection{Category Hit Rate (C-H@K)}
This metric measures the proportion of categories that match users' preferred categories in the top K results of the recommendation list. Here, $\mathcal{U}$ represents the set of users, $K$ is the truncation position, and $\mathcal{C}_u^{\text{pref}}$ denotes the set of historical preferred categories of user $u$ (such as categories that have been clicked or purchased). $hit_u$ is defined as the number of categories in the top K recommendations for user $u$ that belong to $\mathcal{C}_u^{\text{pref}}$. Its calculation formula is:
\begin{equation}
    \text{C-H@K} = \frac{1}{|\mathcal{U}|} \sum_{u \in \mathcal{U}} \frac{hit_u}{K}.
\end{equation}

\subsubsection{Category Normalized Discounted Cumulative Gain (C-N@K)}
This metric evaluates the quality of category ranking in the recommendation list, calculated by the ratio of Discounted Cumulative Gain (DCG) to Ideal Discounted Cumulative Gain (IDCG). Here, $rel(c_i)$ represents the relevance score of the category at the $i$-th position in the recommendation list (e.g., 1 if the user clicks on items of this category, 0 otherwise), and $\log_2(i+1)$ is the position discount factor. The DCG for user $u$ is defined as $\text{DCG}_u = \sum_{i=1}^{K} \frac{rel(c_i)}{\log_2(i+1)}$, and $\text{IDCG}_u$ is the DCG value of the theoretically optimal ranking for user $u$. The global formula is:
\begin{equation}
    \text{C-N@K} = \frac{1}{|\mathcal{U}|} \sum_{u \in \mathcal{U}} \frac{\text{DCG}_u}{\text{IDCG}_u}.
\end{equation}

\subsubsection{Novel Category Proportion (NCP@K)}
This metric calculates the proportion of new categories (that users have never been exposed to) in the top K recommended categories relative to the total categories in the system. Here, $\mathcal{C}$ is the set of all categories in the system, and $\mathcal{C}_u^{\text{hist}}$ is the set of historically clicked categories by user $u$. $new_u$ is defined as the number of categories in the top K recommended categories for user $u$ that do not belong to $\mathcal{C}_u^{\text{hist}}$ (i.e., $\text{new}_u = \sum_{c \in \mathcal{C}_u^K} \mathbb{I}[c \notin \mathcal{C}_u^{\text{hist}}]$). The global formula is:
\begin{equation}
    \text{NCP@K} = \frac{1}{|\mathcal{U}|} \sum_{u \in \mathcal{U}} \frac{\text{new}_u}{|\mathcal{C}|}.
\end{equation}

\subsubsection{Category Long-Tail Proportion (CLTP@K)}

This metric measures the coverage of long-tail categories (categories in the bottom 20\% of click volume ranking) in the top K recommended categories. The set of long-tail categories is defined as $\mathcal{C}^{\text{tail}} \subset \mathcal{C}$, and $tail_u$ represents the number of categories in the top K recommended categories for user $u$ that belong to $\mathcal{C}^{\text{tail}}$ (i.e., $\text{tail}_u = \sum_{c \in \mathcal{C}_u^K} \mathbb{I}[c \in \mathcal{C}^{\text{tail}}]$). The global formula is:
\begin{equation}
    \text{CLTP@K} = \frac{1}{|\mathcal{U}|} \sum_{u \in \mathcal{U}} \frac{\text{tail}_u}{|\mathcal{C}|}.
\end{equation}

\subsubsection{Gross Transaction Value (GTV)}
This metric counts the total transaction amount directly generated by the recommendation system during the experiment. $\mathcal{T}$ is defined as the set of orders guided by recommendations, and $\text{value}(\tau)$ represents the transaction amount of a single order $\tau$. Its calculation formula is:
\begin{equation}
    \text{GTV} = \sum_{\tau \in \mathcal{T}} \text{value}(\tau).
\end{equation}

\subsubsection{7-Day Novel Item Exposure Proportion (7D-NIEP)}
This metric measures the proportion of completely new items (that were not exposed to the user in the past 7 days) among the items exposed on the current day. $\mathcal{I}_u^t$ is defined as the set of items exposed to user $u$ on date $t$, and $\mathcal{E}_u^{[t-7, t-1]}$ represents the set of historically exposed items of the user in the 7 days before day $t$. $new_u$ is the number of newly exposed items on the current day (i.e., $\text{new}_u = \sum_{i \in \mathcal{I}_u^t} \mathbb{I}[i \notin \mathcal{E}_u^{[t-7, t-1]}]$), and $\text{total}_u = |\mathcal{I}_u^t|$ is the total number of exposures. The global formula is:
\begin{equation}
    \text{7D-NIEP} = \frac{1}{|\mathcal{U}|} \sum_{u \in \mathcal{U}} \frac{\text{new}_u}{\text{total}_u}.
\end{equation}

\subsection{Baselines}
\label{Appendix:baselines}
To verify the effectiveness of our method, we compare it with representative baseline models, including 4 exploitation-based baselines: Two-tower \cite{yang2020mixed}, GRU4Rec \cite{hidasi2015session}, SASRec \cite{kang2018self}, and Bert4Rec \cite{sun2019bert4rec}, as well as 7 exploration-based baselines, namely HUCB \cite{song2022show}, pHUCB \cite{song2022show}, NLB \cite{su2024long}, LLM-KERec \cite{zhao2024breaking}, EXPLORE \cite{coppolillo2024relevance}, Google-v1 \cite{wang2025user}, and Google-v2 \cite{wang2025serendipitous}.
\begin{itemize}
    \item \textbf{Two-tower} \cite{yang2020mixed}: A method using the DNN two-tower architecture has replaced the previous MLP method.
    \item \textbf{GRU4Rec} \cite{hidasi2015session}: An RNN-based model that uses GRU to model user behavior sequences.
    \item \textbf{SASRec} \cite{kang2018self}: The first transformer-based sequential recommendation model with unidirectional causal self-attention.
    \item \textbf{Bert4Rec} \cite{sun2019bert4rec}: A representative sequential recommendation model, which employs the deep bidirectional selfattention to model user behavior sequences.
    \item \textbf{HUCB} \cite{song2022show}: A method that constructs a hierarchical item tree, converts exploration into path decision-making from root to leaf nodes, and backpropagates feedback.
    \item \textbf{pHUCB} \cite{song2022show}: A method based on HUCB that avoids upper-level node misleading by expanding qualified visible nodes.
    \item \textbf{NLB} \cite{su2024long}: A scalable exploration algorithm suitable for deep learning-based recommendation systems.
    \item \textbf{LLM-KERec} \cite{zhao2024breaking}: A complementary knowledge-enhanced recommendation system based on LLMs.
    \item \textbf{EXPLORE} \cite{coppolillo2024relevance}: A method that integrates relevance and diversity, and forms a novel recommendation strategy.
    \item \textbf{Google-v1} \cite{wang2025user}: An approach that combines hierarchical planning with LLM inference scaling and distinguishes between the goals of novelty and user alignment.
    \item \textbf{Google-v2} \cite{wang2025serendipitous}: A framework that introduces multimodal large models based on Google-v1 and discovers users' new interests through the chain-of-thought strategy.
\end{itemize}

\subsection{Implementation Details}
\label{appendix:detail}
The sequence filtering click threshold is set to $\tau=5$; the CSA is configured with an embedding dimension of $d=128$, a layer depth of $L=4$, and 4 attention heads. For the Movielens-1M dataset, the configuration of RQ-VAE is as follows: the codebook size is 16, the number of layers of the residual quantizer is $l=4$, and the number of user groups $M_N$ is 50. For the MTRec dataset, the configuration of RQ-VAE is: the codebook size is 64, the number of layers of the residual quantizer is $l=4$, and the number of user groups $M_N$ is 4000. The number of group representative users is $k=8$, the DPO temperature coefficient is $\beta=0.1$, and the alignment threshold is $\tau_{\text{align}}=0.5$. The coefficient of $\alpha=0.4$ of KL divergence. The online model implements daily incremental fine-tuning and updates the offline database, while the offline model is configured differently based on datasets: 30 rounds for Movielens-1M and 1000 rounds for MTRec (each round consists of 5 batches with a batch size of $1024$), with 1\% of $\mathcal{D}_{\text{inc}}$ fed back to the \textit{Novelty LLM} per epoch to synchronously optimize the \textit{Relevance LLM}.

\end{document}